\documentclass[prx,twocolumn,showkeys,showpacs,superscriptaddress,nofootinbib,10pt,floatfix]{revtex4-1}
\usepackage{graphicx}  
\usepackage{subfigure}
\usepackage{amsmath}
\usepackage{verbatim}
\usepackage{dcolumn}
\usepackage{epsf}
\usepackage{colortbl} 
\usepackage{amsthm}
\usepackage{epsfig}
\usepackage{lipsum}
\usepackage{amsmath,amssymb,amsfonts,mathtools,bbm,MnSymbol,mathrsfs,xfrac}   
\usepackage{longtable} 
\usepackage{tabularx}
\usepackage[breaklinks=true,colorlinks=true,linkcolor=blue,urlcolor=blue,citecolor=blue]{hyperref}
\usepackage{comment}
\usepackage{color}
\usepackage[makeroom]{cancel}
\usepackage{ifthen}
\usepackage{amssymb}
\usepackage{soul}

\pdfoutput=1

\renewcommand{\[}{\begin{equation}}
\renewcommand{\]}{\end{equation}}

\newcommand{\ket}[1]{|#1\rangle}
\newcommand{\bra}[1]{\langle#1|}

\newcommand{\mean}[1]{\langle#1\rangle}
\newcommand{\abs}[1]{|#1|}
\newcommand{\ov}[1]{\overline{#1}}
\newcommand{\tr}{\mathrm{tr}}

\newcommand{\R}{{\hat{\rho}}}

\newcommand{\C}{{\mathcal{C}}}
\renewcommand{\P}{\hat{P}}
\newcommand{\bi}{{\boldsymbol{i}}}

\newcommand{\HS}{\mathcal{H}}

\newcommand{\ent}{{\mathrm{ent}}}

\newcommand{\ave}{{\mathrm{ave}}}
\newcommand{\xE}{{\mathrm{xE}}}
\newcommand{\loc}{{\mathrm{loc}}}
\newcommand{\na}{{n_{\!A}}}
\newboolean{showcomments}
\setboolean{showcomments}{True}

\newcommand{\ana}[1]{\ifthenelse{\boolean{showcomments}}{\textcolor{cyan}{[{\bf AA}: #1]}}{}}
\newcommand{\jd}[1]{\ifthenelse{\boolean{showcomments}}{\textcolor{green}{[{\bf JD}: #1]}}{}}
\newcommand{\ds}[1]{\ifthenelse{\boolean{showcomments}}{\textcolor{red}{[{\bf DS}: #1]}}{}}
\definecolor{dfcol}{cmyk}{1, 0.2108, 0.13, 0.3}
\newcommand{\df}[1]{\ifthenelse{\boolean{showcomments}}{\textcolor{dfcol}{[{\bf DF}: #1]}}{}}
\definecolor{mygray}{gray}{0.6}

\theoremstyle{definition}

\usepackage{hyperref}
\begin{document}
\title{Typical and extreme entropies of long-lived isolated quantum systems}
\author{Dana Faiez}
\email{dfaiez@ucsc.edu}
\affiliation{Department of Physics, University of California, Santa Cruz, CA 95064, USA}
\author{Dominik \v{S}afr\'{a}nek}
\email{dsafrane@ucsc.edu}
\affiliation{SCIPP and Department of Physics, University of California, Santa Cruz, CA 95064, USA}
\author{J. M. Deutsch}
\affiliation{Department of Physics, University of California, Santa Cruz, CA 95064, USA}
\author{Anthony Aguirre}
\affiliation{SCIPP and Department of Physics, University of California, Santa Cruz, CA 95064, USA}

\date{\today}

\begin{abstract}

In this paper, we investigate and compare two well-developed definitions of entropy relevant for describing the dynamics of isolated quantum systems: bipartite entanglement entropy and observational entropy. In a model system of interacting particles in a one-dimensional lattice, we numerically solve for the full quantum behavior of the system.  We characterize the fluctuations, and find the maximal, minimal, and typical entropy of each type that the system can eventually attain through its evolution. While both entropies are low for some ``special" configurations and high for more ``generic" ones, there are several fundamental differences in their behavior.  Observational entropy behaves in accord with classical Boltzmann entropy (e.g. equilibrium is a condition of near-maximal entropy and uniformly distributed particles, and minimal entropy is a very compact configuration).  Entanglement entropy is rather different: minimal entropy ``empties out" one partition while maximal entropy apportions the particles between the partitions, and neither is typical.  Beyond these qualitative results, we characterize both entropies and their fluctuations in some detail as they depend on temperature, particle number, and box size. 

\end{abstract}

\maketitle

\section{Introduction}

There are a number of distinct notions of entropy throughout physics -- including for example definitions by Clausius, Boltzmann, Gibbs, von Neumann, Bekenstein-Hawking, and others. Often these notions qualitatively and quantitatively coincide in the limits of large numbers of particles in equilibrium under some set of constraints. Yet, physical systems are commonly in states that are far from equilibrium from the standpoint of fundamental physics.
This raises the question of how these entropies differ out of equilibrium, to what degree their non-equilibrium behavior carries over from the equilibrium case, and how they correspond with our common notions of entropy such as a measure of disorder, an ability to perform work, or of the information content an observer has about the physical system.

In this work, we address these questions, by studying numerically the out-of-equilibrium behavior and the extreme fluctuations of two well-developed notions of entropy that are relevant and interesting in isolated thermodynamic quantum systems.

The first of the two entropies we consider is the entanglement
entropy~\cite{page1993average,grassellientanglement,vidmar2017entanglementquadratic,vidmar2017entanglement}, which is a well-known entropy measure that quantifies the amount of non-local correlation between a subsystem and its compliment. It has a wide range of use and is important in understanding thermalization in isolated systems~\cite{deutsch2013microscopic,zhang2015thermalization,kaufman2016quantum}, quantum correlations and phase transitions~\cite{osborne2002entanglement,latorre2003ground,vidal2003entanglement}, the holographic principle and black hole entropy~\cite{emparan2006black,nishioka2018entanglement}, as well as quantum information theory~\cite{mukohyama1998comments,unanyan2005entanglement,cho2018realistic}. The second entropy we consider is the observational entropy~\cite{safranek2019letter,safranek2019long,safranek2019classical}, which is a generalization of Boltzmann entropy to quantum systems. Originally introduced by von Neumann~\cite{von2010proof,von1955mathematical} as a resolution to the fact that the [von Neumann] entropy does not increase in isolated systems, then briefly mentioned by Wehrl~\cite{wehrl1978general} as ``coarse-grained entropy'', observational entropy has experienced a significant resurgence recently: it was generalized to multiple coarse-grainings~\cite{safranek2019letter,safranek2019long}, found to dynamically describe thermalization of isolated quantum~\cite{safranek2019long,lent2019quantum} and classical~\cite{safranek2019classical} systems, discussed in relationship with other types of entropies~\cite{goldstein2019gibbs}, found to increase under Markovian stochastic maps~\cite{gemmer2014entropy}, and argued for as a natural candidate for entropy production~\cite{strasberg2019entropy} because its definition does not need an explicit temperature dependence.

Fluctuations in entropy were discussed far before these two types of entropy were introduced.  The concept of entropy itself originated from Clausius, who laid the ground work for the second law of thermodynamics in the mid 19th century. It was Boltzmann who interpreted this concept statistically by inventing the well-known \textit{H}-theorem~\cite{boltzmann2003further}, which then led to a new definition of entropy that makes use of the statistical weight of the macrostate; for a given macrostate, the Boltzmann entropy is defined as $S_B=\ln \Omega$, where $\Omega$ is the number of constituent microstates. If the macrostate is an energy macrostate, this entropy is equal to thermodynamic entropy of the microcanonical ensemble~\cite{dunkel2014consistent,hilbert2014thermodynamic,schneider2014comment,buonsante2016dispute}, and is proportional to Clausius's entropy for systems in thermal equilibrium. Considering general (not necessarily energy) coarse-grainings, Boltzmann entropy is typically time-dependent and able to describe systems out of equilibrium, unlike the original definition of entropy~\cite{goldstein2004boltzmann}.\footnote{Note that when we make comparisons with the Boltzmann entropy, we actually mean the definition with general coarse-graining, not with energy coarse-graining which gives what is known as surface, or microcanonical entropy, which stays constant in an isolated system.}

Boltzmann postulated that his entropy (the negative of the quantity $H$) always increases, and did not mention anything about possible downward fluctuations. This was criticizes by Zermelo, and  Boltzmann explains in a later letter~\cite{brush2004history} that fluctuations in entropy are indeed 
unlikely but possible. For example, particles can in principle spontaneously contract into a small space (e.g., corner of a room), and correspond to a macrostate with lower (Boltzmann) entropy. This laid the groundwork for the study of fluctuations in entropy.

Much later, the relations that constrain the probability distribution of entropy fluctuations, i.e. the Fluctuation Theorems (FTs), became one of the most significant discoveries in non-equilibrium statistical physics~\cite{evans1993probability,jarzynski1997nonequilibrium,crooks1999entropy,seifert2012stochastic,luposchainsky2013strong}. Fluctuation relations for closed~\cite{talkner2008microcanonical,esposito2009nonequilibrium,rana2013generalized} and open systems~\cite{esposito2006fluctuation,brunelli2018experimental,manzano2018quantum} pertain when an external force drives the system out of equilibrium. 

These studies do not, however, explore how high or low the entropy of a quantum system can get if it has access to long time scales; this is the focus of  this work. We do this for an isolated system, meaning that 
there is no exchange of energy or particles between the system and the surrounding, and the system evolves unitarily in the absence of any external drive. We also examine what the states with such extreme entropies looks like, how they compare for different types of entropies, and how they depend on system size and inverse temperature.

For generic (i.e. non-integrable) systems and at high temperature, entanglement entropy is equal to the thermodynamic entropy of a subsystem when the full system is in thermal equilibrium~\cite{popescu2006entanglement,deutsch2010thermodynamic,deutsch2013microscopic,santos2012weak,kaufman2016quantum} (with some corrections depending on the size of the system and strength of the interaction term).  One might be tempted to infer that this entropy behaves similarly to Boltzmann entropy, even far from equilibrium. However the studies here of the extreme cases, show that entanglement entropy behaves very differently from that of Boltzmann entropy.

For example, there are macrostates with very many microstates that correspond to minimal entanglement entropy, and macrostates with very few microstates that correspond to maximal entanglement entropy. This shows that outside of equilibrium, entanglement entropy is fundamentally different from Boltzmann's idea of entropy. A type of observational entropy, $S_{\xE}$, on the other hand, associates larger entropy with larger macrostates, in accordance with Boltzmann.

The paper is structured as follows.
In section~\ref{sec:Preliminaries}, we introduce the model at hand as well as the entropies under study.
Next~\ref{sec:distribution}, we examine the probability distribution of entropies over long time unitary evolution of the system and find the minimal and maximal values of entropy, given infinite time. We then compare the states with minimal, maximal and average entropy. In sections~\ref{sec:dependence} and~\ref{sec:dependence2}, we investigate the dependence of extreme values of entropy on system size and inverse temperature, respectively. We find that observational entropy never reaches values significantly below $1/2$ of its maximum value, as argued in a previous study~\cite{deutsch2020probabilistic}; this is in contrast to entanglement entropy of the small subsystem, which can reach values very close to zero in the limit of large system and bath size.
In section~\ref{sec:Maximization of probability}, we provide numerical evidence that the result of~\cite{deutsch2020probabilistic} is correct in the case of a physical system such as a fermionic lattice. Finally, in section~\ref{sec:Connection}, we connect the results of~\ref{sec:dependence} and~\ref{sec:Maximization of probability}: we show that for a highly localized state -- i.e. a state for which the probability of localization in a small region is maximized -- has minimal  observational entropy, but not entanglement entropy. A short introduction to observational entropy can be found in the appendix~\ref{app:oe}.

\section{Preliminaries} \label{sec:Preliminaries}

In this paper we consider a system of $N_p$ spin-less fermions in a 1-dimensional lattice of size $L$ with hard wall boundary conditions. The Hamiltonian describing fermions in $L$ sites is
\[
\begin{split}
\hat{H} = \sum_{i=1}^{L}
[-t({f}_{i}^{\dagger}{f}_{i+1}+h.c.)+V{n}_{i}^{f}{n}_{i+1}^{f}\\
-t'({f}_{i}^{\dagger}{f}_{i+2}+h.c.)+V'{n}_{i}^{f}{n}_{i+2}^{f}].
\end{split}
\]
\vspace{-0.7cm}
\begin{figure}[t]
\centering
\includegraphics[width=0.95\linewidth]{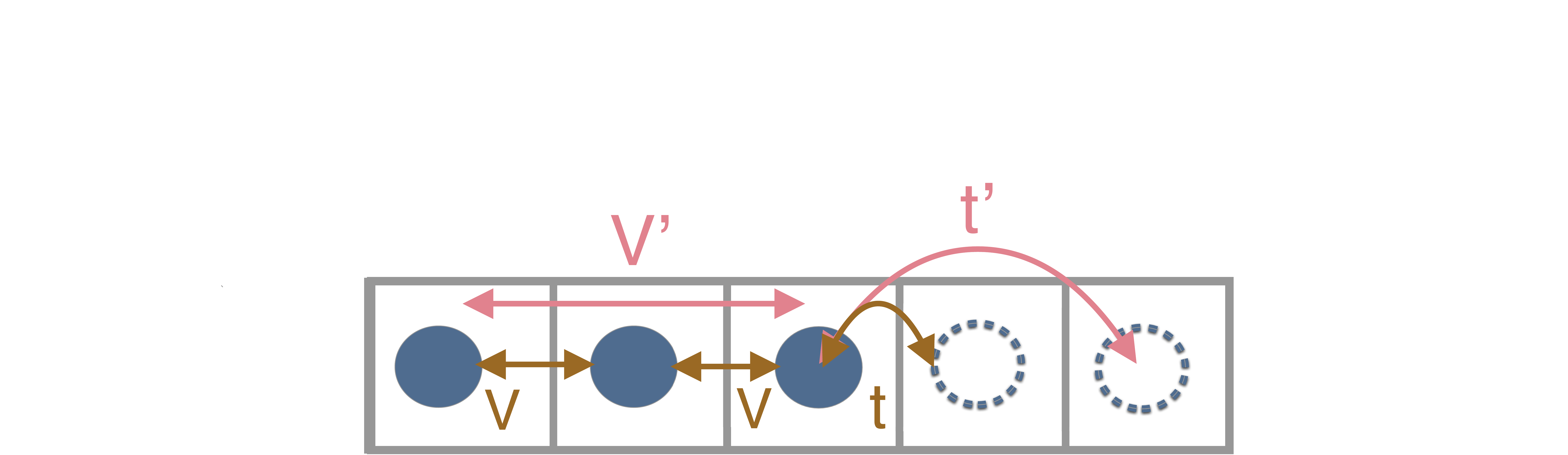}
\caption
{A lattice of size $5$ sites and $3$ particles is shown. The right hand side of the figure illustrates the hopping terms $t$ and $t'$, i.e., particles move to the nearest-neighbor (NN) and next-nearest-neighbor (NNN) sites respectively. The left hand side of the figure shows the interactions of strengths $V$ and $V'$ between NN and NNN respectively.}
\label{lattice}
\end{figure}
\vspace{0.75cm}
(Due to hard-wall boundary conditions, terms with $f_{L+1}$, $n_{L+1}^f$, $f_{L+2}$, and $n_{L+2}^f$ are not included.) Here $f_{i}$ and $f_{i}^{\dagger}$ are fermionic annihilation and creation operators for site $i$ and ${n}_{i}^{f} = {f}_{i}^{\dagger}{f}_{i}$ is the local density operator. The nearest-neighbor (NN) and next-nearest-neighbor (NNN) hopping terms are respectively $t$ and $t'$ and the interaction strengths are $V$ and $V'$ as illustrated in Fig.~\ref{lattice}.

In all simulations, we take $t=t'=1.9$, $V=V'=0.5$. In most simulations we take the inverse temperature to be $\beta=1/T=0.01$ (the reason for this choice is discussed in detail in section~\ref{sec:Maximization of probability}) with exceptions in Figures~\ref{fig.temperature_sent} and~\ref{fig.temperature_sxe}, where we illustrate the dependencies on temperature. We take number of particles $N_{p}$ to be either $2$ or $3$, and we use different system sizes $L$. 
The eigenvalues and eigenvectors of relevant Hamiltonians are computed using exact diagonalization. Using this method however limits us to small size systems due to the exponential rise in computation time and memory requirements with system size, hence the use of only $2$ or $3$ particle systems.

For entanglement entropy, we consider a bipartite system with Hilbert space $\HS_{AB} = \HS_{A} \otimes \HS_{B}$, where A and B label the two partitions. Entanglement entropy is then defined as 
\[\label{Sent}
S_{\mathrm{ent}}(\R_{AB})=-\tr[\R_A\ln\R_A]
\]

where $\R_A=\tr_B[\R_{AB}]$ is the reduced density matrix. This entropy measures the amount of correlations, or ``entanglement'' between $A$, the subsystem of interest, and $B$, the bath. We take sizes of $A$ and $B$ to be $\Delta x=4$ and $L-\Delta x=L-4$ sites respectively. The exception to this is in Sec.~\ref{sec:Maximization of probability} where we consider the smaller subsystem to be of size $\Delta x=5$ sites in order for the subsystem to be large enough to contain all $N_p=3$ particles. 

It worth mentioning that there are other definitions of entanglement entropy in which the system has multiple partitions and the entanglement entropy of the system is the sum of the entanglement entropies of each partition, i.e., the entanglement of each partition with the rest of the system~\cite{coffman2000distributed,rigolin2006operational,deutsch2013microscopic}. However, entanglement entropy is most commonly used in the context of bipartite systems in the literature, for example in relation to the Bekenstein-Hawking entropy of a black hole where the existence of a horizon leads to the bipartition of the degrees of freedom on a Cauchy surface~\cite{das2008black}. It is also common to use in studying quantum information protocols~\cite{bennett1993teleporting,helwig2012absolute} and understanding phases of matter~\cite{zhang2011entanglement,jiang2012identifying}.

Next, we consider observational entropy with position and energy coarse-graining~\cite{safranek2019letter,safranek2019long,safranek2019classical}
\[\label{SxE}
S_{\xE}(\R) \equiv S_{O(\C_{{\hat{X}}},\C_{\hat{H}})}(\R).
\]
The positional (configuration) coarse-graining $\C_{{\hat{X}}}\equiv \C_{\hat{N}_1}\otimes\cdots\otimes\C_{\hat{N}_m}=\{\P_{n_1}\otimes\cdots\otimes \P_{n_m}\}$ defines the partitions of the system (regions), and corresponds to measuring the number of particles in each of the $m$ regions. Energy coarse-graining $\C_{\hat{H}}$ is the coarse-graining given by energy eigenstates of the system, corresponding to measuring the total energy. In contrast with the entanglement entropy, there is not a subsystem or a bath; instead the entire system is divided into equally sized partitions. We set the size of each partition to be $\Delta x=4$, so in a system of size $L$, there will be $m=\frac{L}{\Delta x}=\frac{L}{4}$ number of partitions.

This entropy can be interpreted as ``dynamical'' thermodynamic entropy: it approximates the sum of thermodynamic entropies of each partition~\cite{safranek2019long,safranek2019classical}. As these partitions exchange particles and/or heat, this entropy rises to thermodynamic entropy of the entire system, which corresponds to partitions being in thermal equilibrium with each other. Therefore, $S_{\xE}$ can be loosely interpreted as a measure of how close to thermal equilibrium these partitions are. Precisely, it is an entropy that an observer would associate to a system where $m$ partitions are allowed to exchange energy but not particles, in the long-time limit. In our paper the particles do indeed exchange between the partitions, which is the reason why this quantity is time-dependent. For a short introduction into the framework of observational entropy, see Appendix~\ref{app:oe}.

In all cases, we take the initial state to be a \emph{random pure thermal state} (RPTS) (also known as thermal pure quantum or canonical thermal pure quantum state~\cite{sugiura2012thermal,sugiura2013canonical,nakagawa2018universality}), which we define as
\[
\ket{\psi}=\frac{1}{\sqrt{Z}}\sum_E c_E e^{-\beta E/2}\ket{E},
\] 
where $\ket{E}$'s are the eigenstates of the total Hamiltonian, computed using exact diagonalization. The coefficients $\{c_E\}$ are random complex or real numbers, $c_E\equiv (x_E+iy_E)/\sqrt{2}$, and $c_E\equiv (x_E+y_E)/\sqrt{2}$ respectively, which leads to what will refer to as the \emph{complex} or the \emph{real} RPTS, with $x_E$ and $y_E$ obeying the standard normal distribution $\mathcal{N}(0,1)$, and $Z=\sum_E \abs{c_E}^2 e^{-\beta E}$ is the normalization constant. These states emulate a thermal state, while being pure.  They are then evolved as $\ket{\psi_t}=e^{-i \hat{H}t}\ket{\psi}$.

We would like to point out that in the high-temperature limit, the observational entropy $S_\xE$ of such a state is at all times smaller than the canonical entropy 
$S_{th} \equiv -\tr[\R_{th}\ln\R_{th}]$, where $\R_{th} = \frac{e^{-\beta\hat{H}}}{Z}$ and $Z=\tr e^{-\beta\hat{H}}$ is the partition function. But in the long time limit and for this type of initial state, $S_\xE$ will grow to a value that is very similar to $S_{th}$.
(See Appendix~\ref{app:oe} and Ref.~\cite{safranek2019long} for more detail; this behavior will be illustrated in Figs.~\ref{max-ave-min-Sxe-L} and~\ref{fig.temperature_sxe}.)

\section{Distribution of fluctuations in entropy}\label{sec:distribution}

In this section we explore downward and upward fluctuations in entanglement and observational entropy, and the states achieving extreme values in entropy.

\begin{figure*}
\centering
\begin{minipage}[t]{.45\textwidth}
\includegraphics[width=\columnwidth]{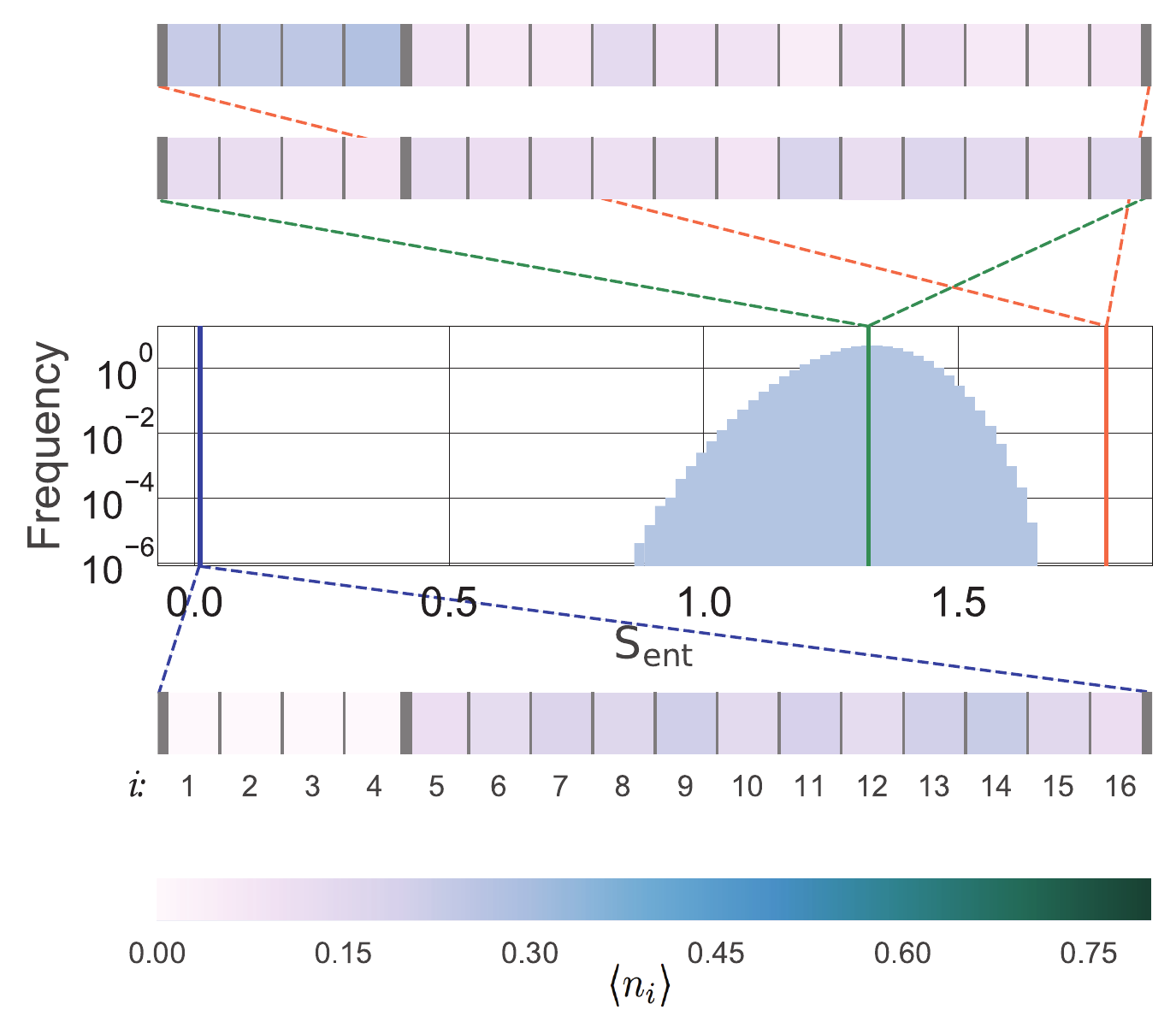}
\caption{Semi-log probability histogram of entanglement entropy, $S_{\ent}$. The y-axis represents the probability of finding the state at any given value of the entropy represented on the x-axis. The heat maps display the particle density ($\mean{n_i}$, $i=1,...L$) on the lattice sites. The left tail of the histogram, representing the downward fluctuations in entropy, can be fitted with a linear function: this shows that fluctuating to small values is exponentially suppressed in this data set. The blue vertical line on the left is the minimum value the entanglement entropy can achieve, and is found using a minimization algorithm. This value is very close to zero. The heat map below shows the particle density on the lattice of the state that corresponds to this minimum. We can see that in this situation, the particles moved almost entirely into the bath, thus naturally producing a separable state. The orange vertical line on the right is the maximum value of the entanglement entropy, and is also obtained by the minimization algorithm. The heat map above shows the particle density on the lattice of the state that corresponds to this maximum. In this situation, both the subsystem and the bath have the same number of particles, hence we see a higher density of particles in the subsystem. The state that gives the maximal entanglement entropy, is very far from the thermal equilibrium state.}
\label{Hist_Sent}
\end{minipage}\qquad
\begin{minipage}[t]{.45\textwidth}
\includegraphics[width=\columnwidth]{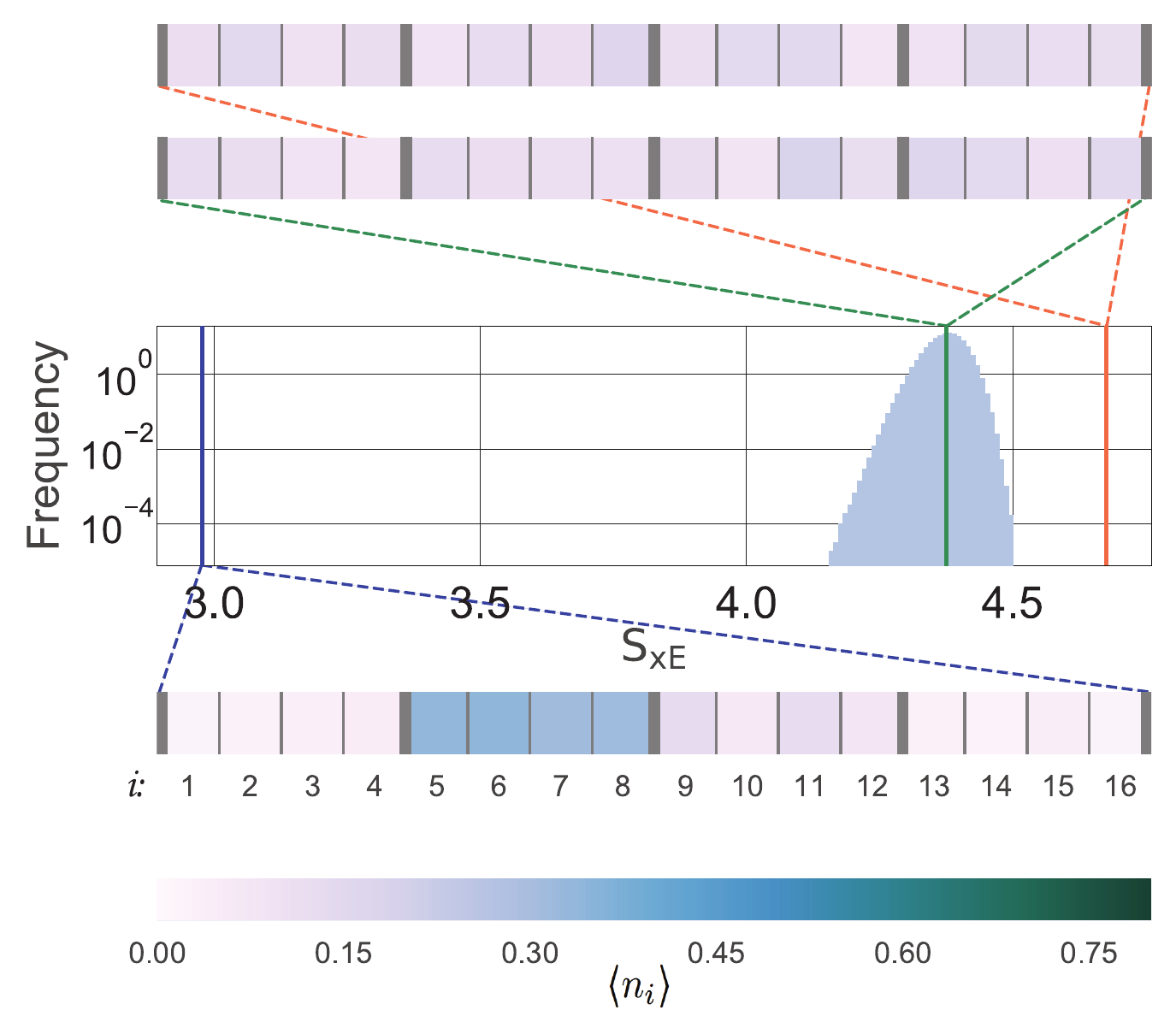}
\caption{Similar to entanglement entropy, downward fluctuations of $S_{\xE}$ to small values is exponentially suppressed in this data set. However, in contrast to entanglement entropy, the minimum of $S_{\xE}$ represented by the blue vertical line on the left does not go to zero; it is at about $63\%$ of the maximum value. This is because it is impossible to localize the particles entirely into the small region, and the remaining regions still contribute significantly to the total entropy. As one can see from the heat map of the state corresponding to the minimum, a significant number of particles moved into one of the partitions of size $4$ sites, resulting in partitions being far from thermal equilibrium from each other. The vertical lines represent the minimum, the average, and the maximum of $S_{\xE}$ from left to right. The heat map above shows the uniform distribution of particles for such state. In contrast with entanglement entropy, the states that give the average and maximal values of $S_{\xE}$ are very similar to each other, as one would expect from the behavior of Boltzmann entropy.}
\label{Hist_Sxe}
\end{minipage}
\end{figure*}

First, we plot histogram of fluctuations in entanglement (Fig.~\ref{Hist_Sent}) and observational entropy (Fig.~\ref{Hist_Sxe}), in a system of size $L=16$: starting from a complex RPTS, the system is evolved, and at each small fixed time step we read out the value of entropy. Evolving for a long time, we therefore achieve sufficient statistics that tells us how likely it is to find any given value of entropy.

We can also ask what the minimum and maximum values of entropy are, given infinite time. Due to the exponential suppression of these extreme values, histogram cannot provide this minimum; we therefore use a minimization algorithm, explained below, and add the results to the histogram (orange and blue vertical lines in Figs.~\ref{Hist_Sent} and~\ref{Hist_Sxe}).

To find the extreme values of entropy we use the simplex search algorithm~\cite{nelder1965simplex}. For a given $L$ and $\beta$, we initialize the state in the same complex RPTS as the one we used to create the histograms in~\ref{Hist_Sent} and~\ref{Hist_Sxe}. We then find the maxima and minima for this initial state by maximizing over phases $\phi_E=E t$. As long as differences of $E$'s are irrational (or close to being irrational), this method must give the same result as maximizing over all times $t$. 

For each histogram, we provide three heat maps, displaying the particle density ($\mean{n_i}$, $i=1,...L$) on the lattice sites for states of maximal, average, and minimal entropy.

\subsection{Entanglement entropy}
We can see that entanglement entropy achieves a minimal value that is very close to zero. We plot the heat map (below the histogram in Fig.~\ref{Hist_Sent}) of the particle density of the state that corresponds to this minimum. We can see that in this situation, the particles moved almost entirely into the bath, thus naturally producing a separable state $\ket{\psi_{\min}}\approx\ket{0}_A\otimes\ket{\psi}_B$, where $\ket{0}_A$ denotes vacuum in the subsystem. One might think that an alternative state $\ket{\psi_{\mathrm{min}}}\approx\ket{\psi}_A\otimes\ket{0}_B$, could also lead to zero entanglement entropy. However, as it is explained in the Section~\ref{sec:Maximization of probability}, one can not cluster all particles in a small region, when starting in a RPTS.

On the other hand, the state with maximum value of entanglement entropy is the one where the subsystem and the bath contain the same average number of particles. The smaller region therefore has a higher density of particles, as illustrated on the heat map. Intuitively, there have to be some particles in the subsystem and some in the bath, for any correlations to exist; and to create the maximum correlation, there should be the same amount of particles on either side. As can be seen from comparing the heat maps in Fig.~\ref{Hist_Sent}, the state that has the maximum entanglement entropy is quite different from the thermal equilibrium state, where particles are distributed uniformly.

\subsection{observational entropy $S_{\xE}$}
The minimum in $S_{\xE}$ is achieved by simply localizing the particles in one of the regions to the extent possible (it does not matter significantly which one, as they all give almost equal entropy; however, if one of the regions was smaller than the others, it would localize into this smallest region). The minimal value of $S_{\xE}$ never goes below about half of the maximal entropy; this,  again, has to do with the inability to cluster all particles in a small region, when starting in an RPTS (see Sec.~\ref{sec:Connection} for a better intuition). The maximum of $S_{\xE}$ is given by a state where particles are uniformly distribution across all regions. $S_{\xE}$ is therefore in accordance with the Boltzmann entropy, in contrast to entanglement entropy.

\section{Dependence of extreme values on the system size}\label{sec:dependence}

\begin{figure}
\centering
\includegraphics[width=1\linewidth]{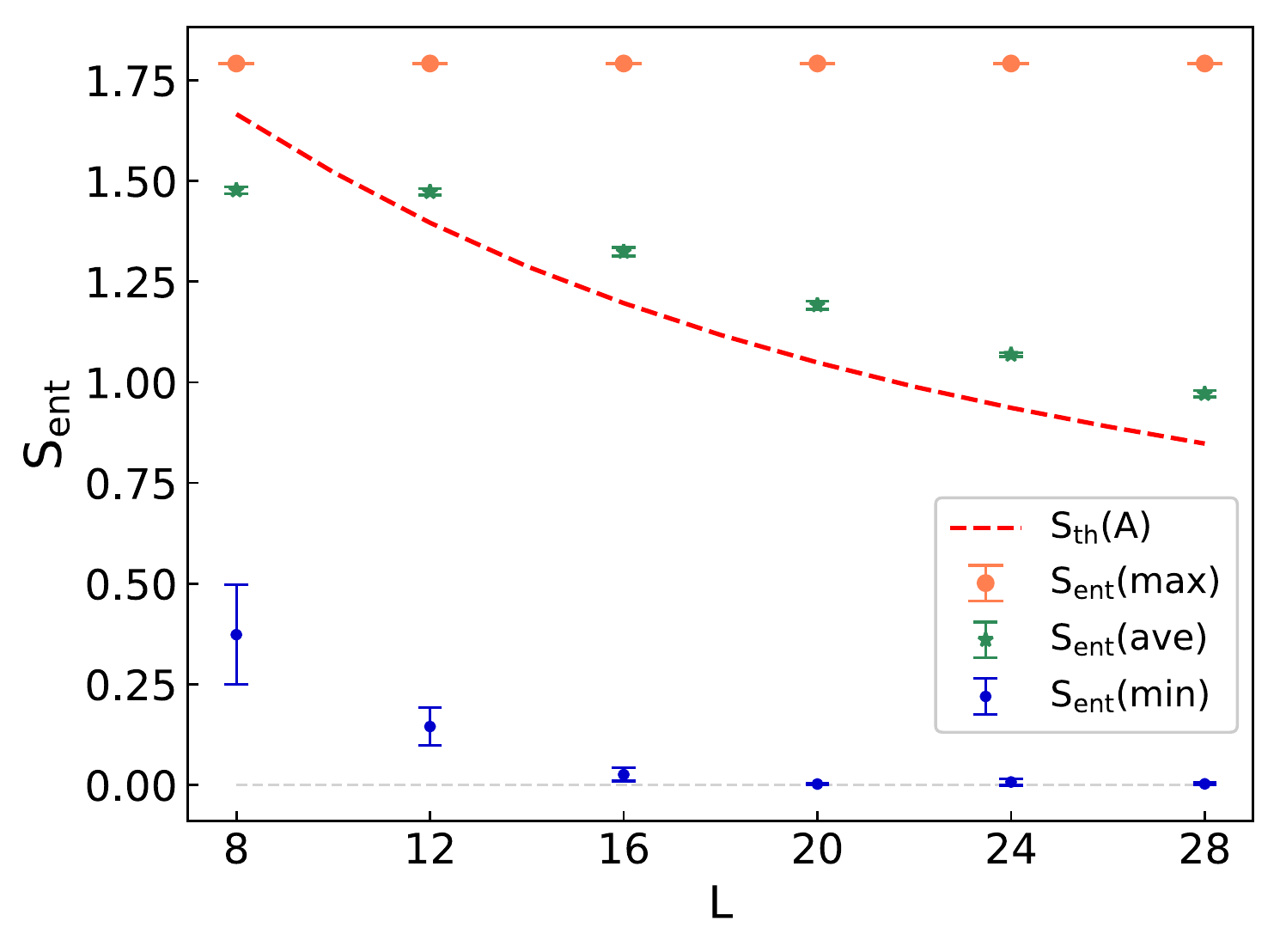}\\
\caption {The minimum (blue dots), maximum (orange circles), and average (green stars) values of entanglement entropy is computed for $6$ different initial random states (complex RPTSs); the mean and standard deviation of these $6$ values are illustrated in this figure for various system sizes.
$S_{\ent}(\min)$ approaches zero in the limit of large $L$: disentanglement of the two regions is mostly done by moving the particles into the bath and emptying the subsystem and in this limit: almost all particles are in the bath and none in the subsystem. Hence $S_{\ent}(\min)$ reaches zero. In contrast $S_{\ent}(\max)$ is independent of $L$ (and hence the size of the bath): maximum entanglement is achieved when particles are equally distributed in each region, and enlarging the bath, given this distribution of particles, does not affect the entanglement entropy of the system. $S_{\ent}(\ave)$ decreases with $L$, and is expected to approach zero for large system sizes as almost all particles on average would be in the bath when the system is large and the subsystem is small. We also plot the thermodynamic entropy~\eqref{eq:SentaveSth} of the subsystem during equilibrium (red dashed line), which is expected to equal $S_{\ent}(\ave)$ in the limit of large system sizes and high temperatures (see Eq.~\eqref{eq:SentaveSth}). Noticeably lower value of $S_{\ent}(\ave)$ (by about $\ln 2$) for $L=8=2\Delta x$ is due to Page curve~\cite{nakagawa2018universality}. We stress that maximal entanglement entropy does not equal the average.
}
\label{max-ave-min-Sent_L}
\end{figure}

\begin{figure}
\centering
\includegraphics[width=1\linewidth]{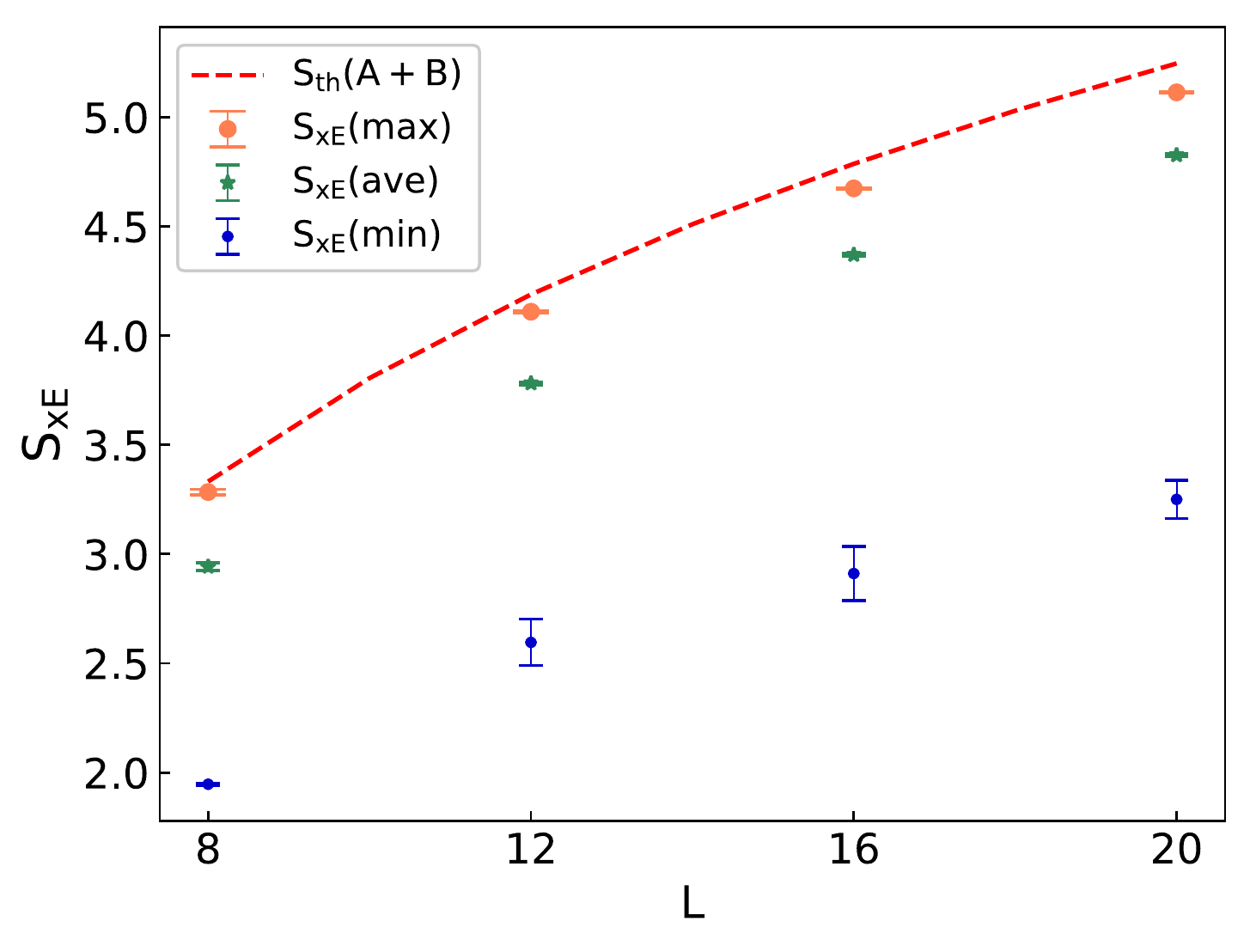}\\
\caption {The minimum (blue dots), maximum (orange circles), and average (green stars) values of observational entropy $S_{\xE}$ is computed for $6$ different initial random states(complex RPTSs); the mean and standard deviation of these $6$ values are illustrated in this figure for various system sizes. Partitions have equal sizes fixed at $\Delta{x}=4$. All $S_{\xE}(\min)$, $S_{\xE}(\mathrm{ave})$, $S_{\xE}(\max)$ increase with $L$, and $S_{\xE}(\mathrm{ave})\approx S_{\xE}(\max)$ are approximately equal the thermodynamic entropy of the full system $S_{th}(A+B)$, as expected from the theory.}
\label{max-ave-min-Sxe-L}
\end{figure}

Next, we study the dependence of the minimum, maximum, and mean values of entropy on the system size. The minimum and maximum values are found using the minimization algorithm as in figures~\ref{Hist_Sent} and~\ref{Hist_Sxe}, and the average value is found by evolving the system for a long time.

\subsection{Entanglement entropy}
These values are shown for entanglement entropy in Fig.~\ref{max-ave-min-Sent_L}. The size of the subsystem is kept fixed at $\Delta x=4$ while the system size (and hence the bath size) is varied such that $8\leq L \leq28$. The number of particles and inverse temperature are kept fixed at $N_p=2$ and $\beta=0.01$ respectively. For each $L$, we initialize the state in $6$ different complex RPTS, and then find the minima, maxima, and average values of entropy for each one of them. We plot the mean value of these six minima, maxima, and average, as well as the standard deviation (denoted as error bars), for a given system size $L$.

We see the decrease in minimum entanglement entropy in Fig.~\ref{max-ave-min-Sent_L} with increasing system size $L$. As we discussed in the previous section in relation with Fig.~\ref{Hist_Sent}, the entanglement between the subsystem and the bath is reduced mostly by moving all the particles into the bath. It is clear that as the bath (of size $L-\Delta x$ where $\Delta x$ is fixed) gets larger, it becomes easier to cluster all the particles in the larger bath, which makes the subsystem emptier, thus creating a state that resembles very closely a product state, and thus has a very small entanglement entropy.

It is important to emphasize that reduction in entanglement entropy is not achieved through disentangling the particles, but by disentangling the regions through the means of particles hopping and emptying the smaller region. 
Therefore the following question is raised: how much entropy would be reduced if particles' hopping between the regions was forbidden?
A simulation of this case -- where the hopping terms between the two regions are zero -- revealed that the reduction of entanglement entropy is much smaller: about a $20\%$ reduction.

The upper bound on maximum of entanglement entropy was derived in~\cite{faiez2020much} for closed, fermionic and bosonic systems. Specifically, in Fig.~\ref{max-ave-min-Sent_L}, where a (1-dimensional) fermionic lattice is considered, 
we have
\[\label{eq:upperbound}
S_{\ent}(\max) \leq \ln{\sum_{n_{A=0}}^{N_p} \min\Bigg\{\binom{\Delta x}{n_A}
,\binom{L-\Delta x}{N_p-n_A}\Bigg\}}.
\]

In the case of $N_p=2$ and $\Delta x=4$ explored in Fig.~\ref{max-ave-min-Sent_L}, $S_{\ent}(\max)$ achieves exactly this upper bound at $\ln 6=1.79$. 
The upper bound~\eqref{eq:upperbound} is independent of the size of the bath in the limit of large $L$, which explains the constant maximum value in Fig.~\ref{max-ave-min-Sent_L} (the large $L$ in this case is already $L\geq 12$, and $L=8$ gives coincidentally the same value).

We should also note that, based on relation \eqref{eq:upperbound} found in~\cite{faiez2020much}, one can see that this upper bound only depends on a few parameters, namely the size of the total system, size of the subsystem, the total number of particles (which is assumed to be fixed), and the assumption that the system is pure. Hence, it does not matter where the subsystem or bath is placed inside the system (for example in the middle or at the edge of the lattice).

The average entanglement entropy in the high temperature limit should be approximately equal to the thermodynamic entropy of the subsystem~\cite{popescu2006entanglement,deutsch2013microscopic,yamaguchi2018theorems}, which is a fraction of the total thermodynamic entropy,
 \[\label{eq:SentaveSth}
S_{\ent}(\ave) \approx S_{th}(A) \approx \frac{\Delta{x}}{L}S_{th}(A+B),
 \]
where $S_{th}$ is computed as the von Neumann entropy of a thermal state, i.e. $S_{th}(A+B) = -\tr[\R_{th}\ln\R_{th}]$ for the full system, where $\R_{th} = \frac{e^{-\beta\hat{H}}}{Z}$ and $Z=\tr e^{-\beta\hat{H}}$ is the partition function, while $S_{th}(A) = -\tr[\R_{th}^{(A)}\ln\R_{th}^{(A)}]$ where $\R_{th}^{(A)}=\frac{e^{-\beta\hat{H}_A}}{Z}$ for the reduced system.
Relation \eqref{eq:SentaveSth} has been confirmed in various numerical simulations~\cite{sorensen2007quantum,zhang2015thermalization,kaufman2016quantum}. We see that this prediction, plotted as a red dashed line in Fig.~\ref{max-ave-min-Sent_L}, fits quite well with the data. (The relation \eqref{eq:SentaveSth} is only approximately true in the high temperature limit and not the low temperature limit, as is discussed in~\cite{hanggi2008finite,horhammer2008information}.) 

Comparing the maximum value of entanglement entropy with the average, we note that $S_{\ent}(\max)$ is constant while $S_{\ent}(\ave)$ decreases with $L$. This is expected, since the average state spreads the particles uniformly over the entire system (creating less entanglement between the subsystem and the bath), while maximizing entanglement entropy maximizes correlations by 
distributing particles throughout the lattice such that the average number of particles in the subsystem follows the necessary but not sufficient condition for the state to be maximally entangled, stated in Ref.~\cite{faiez2020much}. Based on this condition, the mean number of particles in sub-lattice $A$ is,

\[\label{eq:na_maxSent}
\ov{\na}=\sum_{\na=\max\{0,N_p-L+\Delta{x}\}}^{\min\{N_p,\Delta{x}\}} p_{\na}\na\] 

where $p_{\na}=\frac{d_\na}{\sum_{\na=\max\{0,N_p-L+\Delta{x}\}}^{\min\{N_p,\Delta{x}\}} d_\na}$, with
$d_\na=\min\big\{\binom{\Delta{x}}{\na},\binom{L-\Delta{x}}{N_p-\na}\big\}$ is the probability of measuring $\na$ particles in sub-lattice $A$. We should note that in the numerical studies shown here, the average number of particles in the subsystem is equal to that of the bath but this is not in general the case.
This discussion adds to Fig.~\ref{Hist_Sent} in demonstrating the difference between states leading to the average and the maximal entanglement entropy.

\subsection{Observational entropy $S_{\xE}$}

Using the same procedure, we find the mean values of minima, maxima, and averages of $S_{\xE}$, and their variances, and plot them as a function of the system size in Fig.~\ref{max-ave-min-Sxe-L}. Partitions have equal sizes fixed at $\Delta{x}=4$ and the system size $L$ (and therefore the number of partitions $m=\frac{L}{\Delta{x}}$) is varied.

The minimum values of observational entropy $S_{\xE}$ reduces to about a half of its maximum value independent of the system size, as long as it is large. These values could be indirectly estimated by simply assuming that the spatial localization is key in minimizing the entropy (see Fig.~\ref{fig.local_SxeSent} and Eqs.~\eqref{Sxe_Sth} and~\eqref{eq:SxEgtHalfSth}). 

The maximum value of $S_{\xE}$ is almost exactly the same as the thermodynamic entropy of the full system, and very close to the average value of $S_{\xE}$. This is expected from the theory~\cite{safranek2019long}, that shows $S_{\xE}(\mathrm{ave})\leq S_{\xE}(\mathrm{max})\lessapprox S_{th}$, and $S_{\xE}(\mathrm{ave})$ differs from thermodynamic entropy $S_{th}$ by order-1 corrections (that depend on the energy distribution of the initial state), by $\ln N$ corrections (that depend on how close the initial state is to the thermal state), both of which become irrelevant in the thermodynamic limit, and by finite-size corrections (coming from interaction energy between partitions), which become irrelevant when partitions are large enough.

\section{Dependence of extreme values on temperature}\label{sec:dependence2}

\begin{figure}[t!]
\centering
\includegraphics[width=1\linewidth]{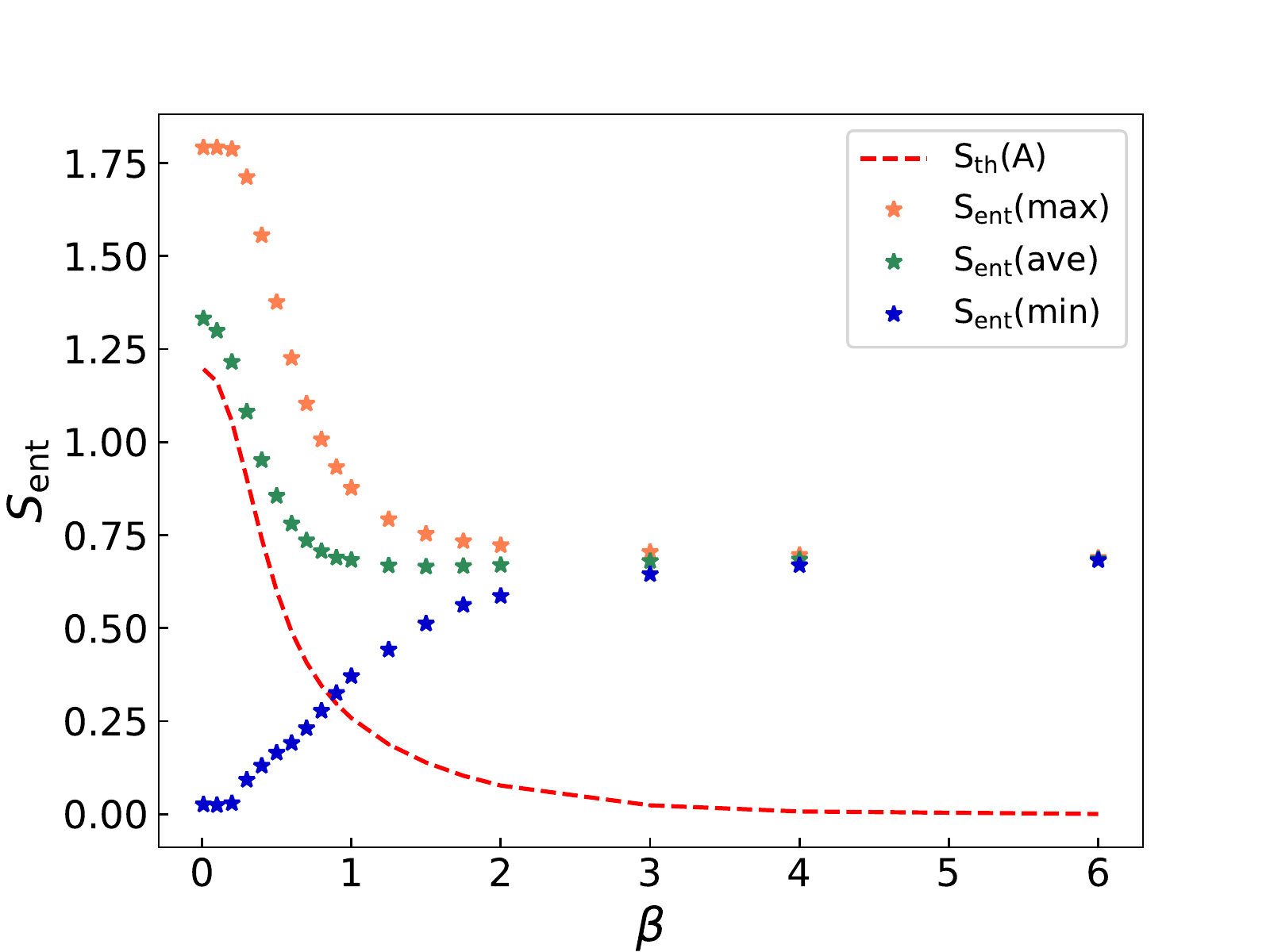}\\
\caption {The minimum (lower blue stars), maximum (orange upper stars), and average (green middle stars) values of entanglement entropy is computed for $6$ different initial random states (complex RPTSs); the means of these $6$ values are illustrated in this figure for various inverse temperatures, $\beta$. We take $L=16$, $\Delta{x}=4$, and $N_p=2$. In low $\beta$ limit,  $S_{\ent}(\mathrm{ave})$ follows the volume law, and is approximately equal the thermodynamic entropy of the subsystem $S_{th}(A)$. In high $\beta$ limit, the initial state is practically the energy ground state, and therefore it does not evolve, so all values coincide, at a value given by the area law.}
\label{fig.temperature_sent}
\end{figure}

In this section, we look at the dependencies of the average and both extremes of $S_{\ent}$ and $S_{\xE}$ on inverse temperature $\beta$. Each data point in Figures~\ref{fig.temperature_sent} and~\ref{fig.temperature_sxe} are computed by taking the mean of the $\min$, $\max$, and average entropies over $6$ different complex RPTSs. We also included the thermodynamic entropy of the subsystem, $S_{\mathrm{th}}(A)$, and of the total system, $S_{\mathrm{th}}(A+B)$,  in Figures~\ref{fig.temperature_sent} and~\ref{fig.temperature_sxe} respectively.

\subsection{Entanglement entropy}

Fig.~\ref{fig.temperature_sent} plots the entanglement entropy versus $\beta$. As one would expect, there are high fluctuations in the low $\beta$ (high temperature) limit. In this limit, the average entanglement entropy coincides with the thermodynamic entropy of the subsystem, which is known as the Volume law~\cite{nakagawa2018universality}. Both maximal and minimal entanglement entropy diverge from the average at low $\beta$, and are almost constant in this limit: $S_{\ent}(\max) \approx 1.79$ (which is the high-temperature limit obtained previously in Fig.~\ref{max-ave-min-Sent_L}), and $S_{\ent}(\min) \approx 0.05$. 
There are almost no fluctuations in the opposite high $\beta$ (low temperature) limit, where the thermal state is almost identical to the ground state, and therefore it does not evolve. The entanglement entropy approaches a constant value given by the Area law~\cite{eisert2008area,laflorencie2016quantum,cho2018realistic}. The difference between entanglement entropy and thermodynamic entropy is discussed in more detail in~\cite{horhammer2008information}.

\subsection{Observational entropy $S_{\xE}$}
\begin{figure}[t!]
\centering
\includegraphics[width=1\linewidth]{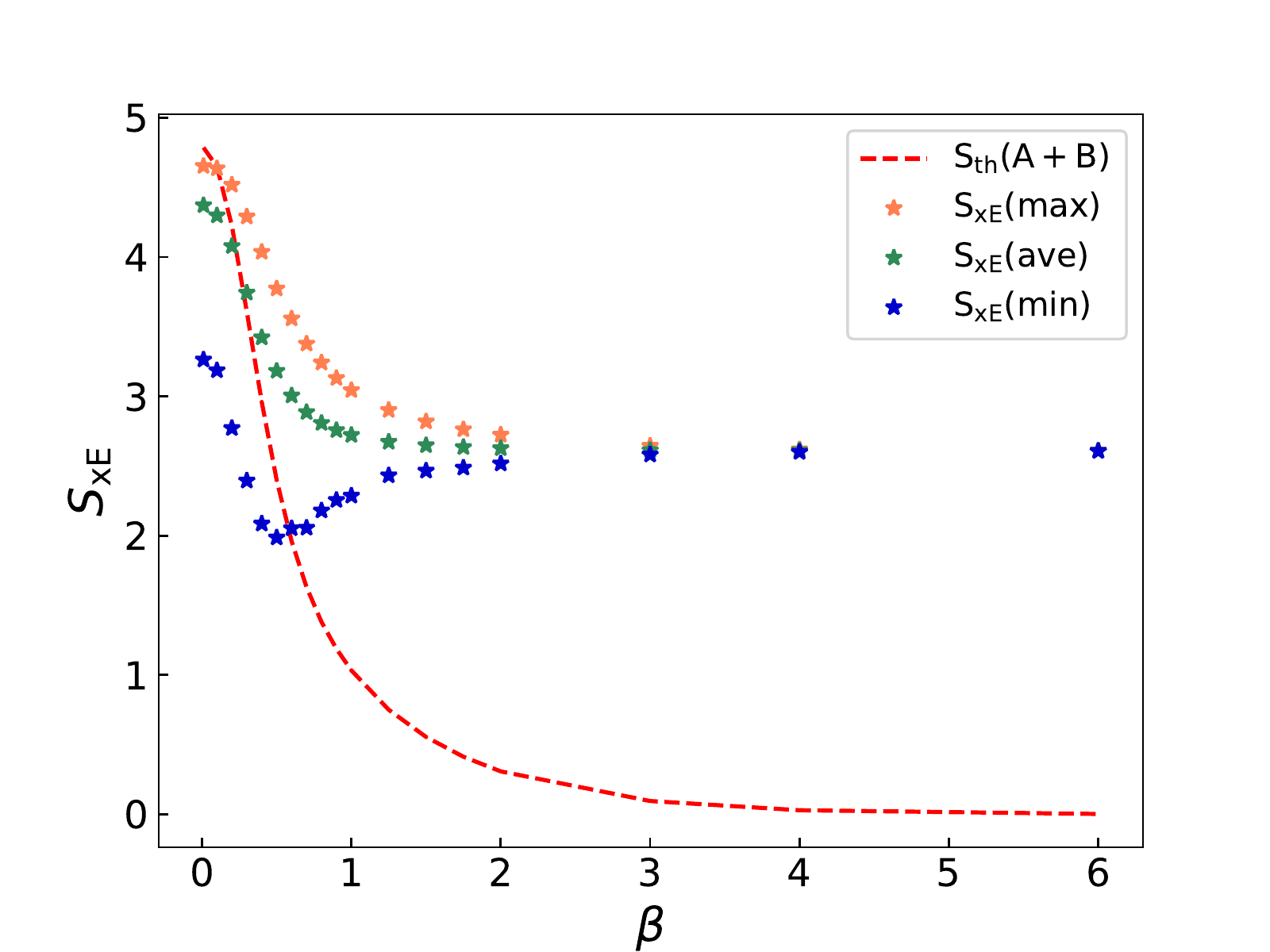}\
\caption {The minimum (blue lower stars), maximum (orange upper stars), and average (green middle stars) values of observational entropy $S_{\xE}$ is computed for $6$ different initial random states (complex RPTSs); the means of these $6$ values are illustrated in this figure for various system sizes. We take $L=16$, $\Delta{x}=4$, and $N_p=2$. In low $\beta$ limit,  $S_{\xE}(\mathrm{ave})\approx S_{\xE}(\max)\approx S_{th}(A+B)$, and  $S_{\xE}(\min)$ has the same shape, and about a half of the maximum value, as expected from Eqs.~\eqref{Sxe_Sth} and \eqref{eq:SxEgtHalfSth}. All values coincide in the high $\beta$ limit where the initial state is practically the energy ground state. Its higher value compared to $S_{th}(A+B)$ is expected from the fact that measuring position of this highly non-local state first, creates a large uncertainty in energy, and therefore also large $S_{\xE}$. The dip in $S_{\xE}(\min)$ is the result of two competing factors: higher temperature results in higher entropy on average, but also higher ability of the system to localize, and therefore possibly lower values of $S_{\xE}$. $\beta\approx 0.5$ is the lowest possible temperature such that the state can localize in one of the bins of size $\Delta x=4$.}
\label{fig.temperature_sxe}
\end{figure}

Fig.~\ref{fig.temperature_sxe}  plots the observational entropy $S_{\xE}$ versus $\beta$, and we took the same settings as with entanglement entropy. One can notice two interesting features in this graph.

First, values of $S_{\xE}$ at high $\beta$ (low temperature) limit are quite large, and do not seem to follow the $S_{th}(A+B)$ anymore. The fact that the $S_{xE}$ is not zero in this low temperature limit is because measuring position does not commute with measuring energy. By measuring the position of the ground state, which is highly non-local, one would add a lot of energy to it, as well as uncertainty in energy. Therefore, since $S_{\xE}$ measures the total uncertainty when measuring the position first and then energy, this total uncertainty will be large. $S_{\xE}$ can be also interpreted as a thermodynamic entropy of the system, as if the numbers of particles in each bin were fixed, but the energy between the bins was still allowed to exchange~\cite{safranek2019classical,safranek2019letter,safranek2019long}. It therefore makes sense that the value of this entropy is relatively large, since by measuring the position we fix the number of particles in each bin, and this state has a relatively large thermodynamic entropy. This effect gets to be smaller ($S_{\xE}$ for high $\beta$ is smaller), when size of the partition $\Delta x$ becomes large compared to the size of the full system, since position measurement does not affect energy as much in that case. We note that this is a purely quantum effect, however, switching the order of coarse-grainings (while taking some small coarse-graining of width $\Delta E$ in energy as well), $S_{Ex}$  leads to an entropy that is bounded above by $S_{th}(A+B)$ even at such low temperatures. (See Appendix~\ref{app:oe} for details.) This is because measuring energy of a ground state does not affect this state at all, and additional measurement in position does not add any new information (see Theorem 8 in~\cite{safranek2019long}). This effect was not noted in the original paper~\cite{safranek2019long}, mainly because defining microcanonical entropy at such low temperatures is problematic, as the energy density of states is not well defined.\footnote{Fig. 7 in~\cite{safranek2019long} does not show $S_{\xE}$ nor microcanonical entropy for really low, or really high energies $E$.}


The second interesting feature of this graph is the dip in $S_{\xE}(\min)$ at $\beta\approx 0.5$. This dip is a result of two competing factors: first, by increasing the temperature, we increase the ability of the system to localize. Generally, localizing the system in one of the partitions leads to a decrease in $S_{\xE}$ (see Section~\ref{sec:Connection}). Thus, with high enough temperature the system is able to localize in one of the partitions of size $\Delta x=4$ and decrease the entropy. However, further increasing temperature does not help in decreasing $S_{\xE}(\min)$ anymore, as the further ability to localize is already below the resolution of the positional coarse-graining in $S_{\xE}$, and its only effect is then an increase in the total thermodynamic entropy, and hence also an increase $S_{\xE}(\min)$.

That is also why we see the increase in $S_{\xE}(\min)$ for really high temperature (low $\beta$), in a shape that approximately follows Eqs.~\eqref{Sxe_Sth} and~\eqref{eq:SxEgtHalfSth}.

\section{Maximal probability of localization}\label{sec:Maximization of probability}

\begin{figure}
\includegraphics[width=0.9\hsize]{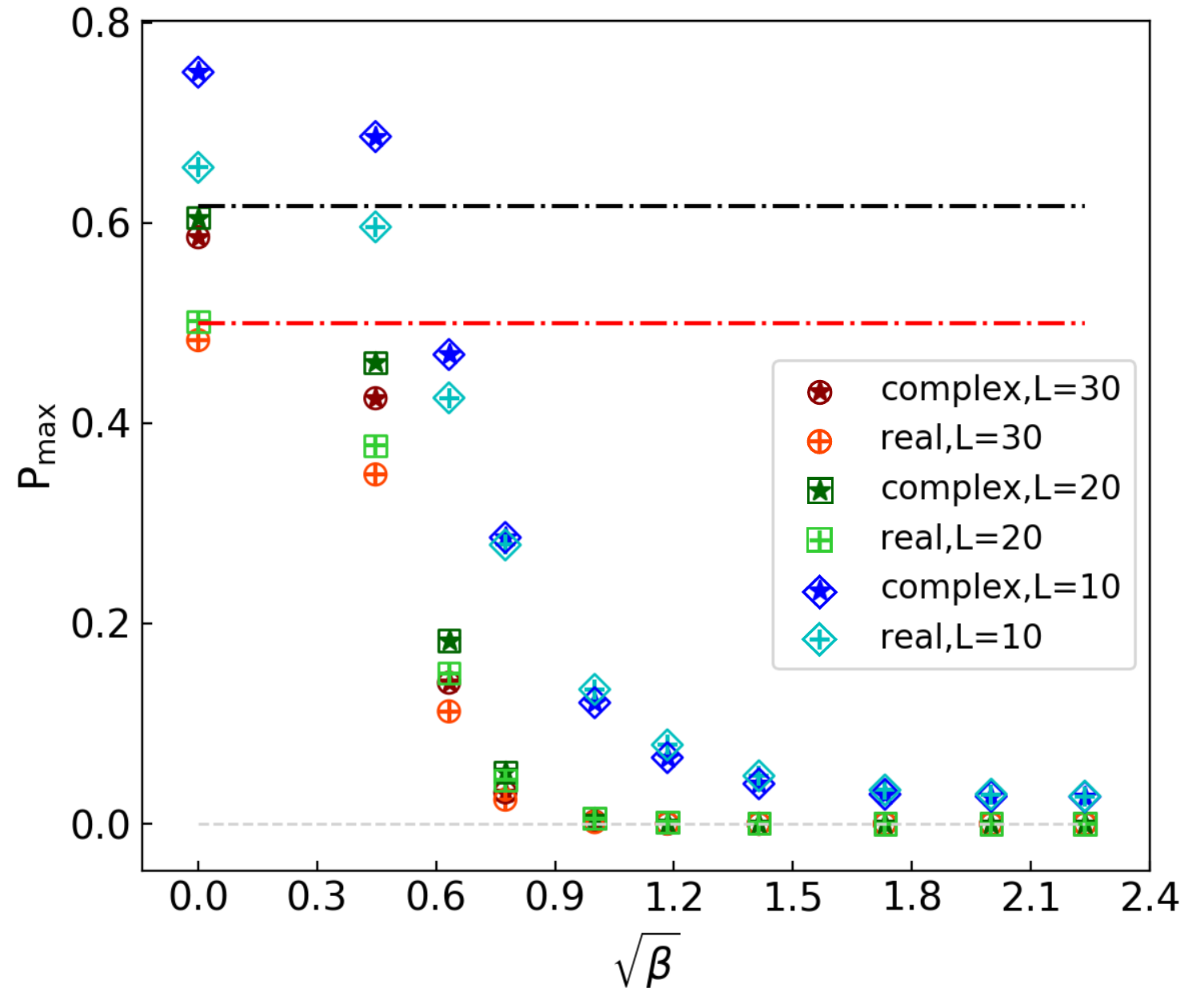}\\
~~~(a)\\
\includegraphics[width=1\hsize]{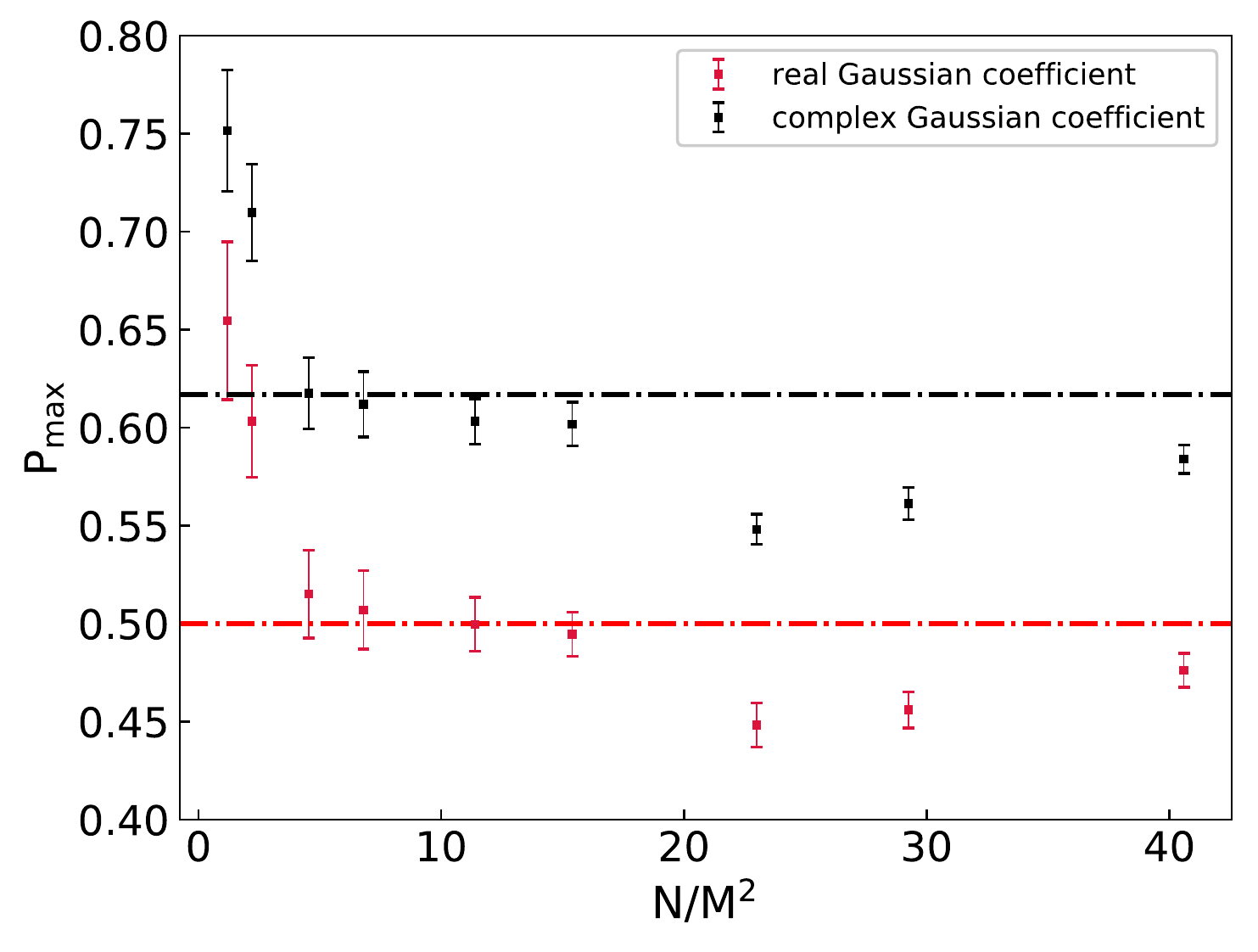}\\
~~~(b)\\
\caption  {(a) Maximum probability $P_{\max}$ of localizing all particles in the middle $5$ sites for real (crosses) and complex (stars) initial RPTSs, in a lattice of size L=10 (blue diamonds), 20 (green squares), and 30 (red circles), with $3$ particles as a function of $\sqrt{\beta}$. This plot illustrates that at low $\beta$, $P_{\max}$ approaches different constant values for real ($0.5$ red lower line) and complex ($\pi^{2}/16$ black upper line) RPTSs when the system size is large enough. In the same limit, $P_{\max}$ approaches unity for smaller systems. For higher values of $\beta$, $P_{\max}$ approaches zero independent of system size. (b) The maximal probability $P_{\max}$ is computed for a range of dimensions of Hilbert space $N$, while $M$ --- dimension of the subspace of the Hilbert space associated with ``all particles in the localized region'' --- is kept fixed. Hence the size of the physical region in which particles are localized in is kept fixed as well, at $\Delta{x}=5$ sites. For each $N$, we start in $100$ different real and complex RPTSs with the same temperature, and plot the mean and standard deviation of $P_{\max}$ (the red lower set of bars as real, and the upper set of black bars as complex). This plot indicates that in the limit of large system sizes, the maximum probability of localization of all particles into a small region approaches $\sim 0.5$ (red lower line) in the case of real initial states and $\sim \pi^{2}/16$ (black upper line) in the case of complex initial states.}
\label{fig.Pmax}
\end{figure}

In this section we show numerically that the result of Deutsch et al.~\cite{deutsch2020probabilistic} -- shown analytically for a toy model with random energy eigenvectors as well as for a non-degenerate weakly
interacting gas -- holds true for a physical system of a fermionic lattice. We do this because in Sec.~\ref{sec:Connection} we would like to use this result to explain the connection, already hinted in the previous sections, between the spatial localization and the minimization of entropies.

In particular, Deutsch et al.~showed that starting from a RPTS, under certain conditions, the maximum probability $P_{\max}\equiv P_{(N,0)}$ that all particles are localized into the subsystem of interest is $1/2$ in the case of initial real RPTS and $\pi^{2}/16$ in the case of complex RPTS. This, as shown in section~\ref{sec:Connection}, is key in minimizing $S_{\xE}$.

We are going to require that the same conditions as in~\cite{deutsch2020probabilistic} to be satisfied: The first condition is that the dimension $M$ of the subspace $X$ (the subspace of Hilbert space associated with ``all particles being in the subsystem of interest'') is much smaller than the dimension $N$ of the full system, $N\gg M^2$, which can be for example satisfied in the case of dilute gas (small number of particles) when the size of the subsystem of interest $\Delta x$ into which we localize the particles is much smaller than the size $L$ of the full system, $L\gg\Delta x$. At the same time, the second condition is that the size of the subsystem is much larger than a thermal wavelength (specified below), $\Delta x\gg \lambda_T$. The third condition is that the size of the subsystem of interest, $\Delta x$, is also much greater than the scattering length, i.e., we consider the Hamiltonian with only local interactions, leading to a weakly correlated system. However unlike what is used in~\cite{deutsch2020probabilistic} -- in which the energy eigenstates of the toy model are randomly distributed or are that of a non-degenerate weakly interacting gas -- in our case the energy eigenstates are that of a Hamiltonian modeling a fermionic lattice.

First, we investigate the second condition, $\Delta x\gg \lambda_T$ in more detail. At any value of $\beta$, there exists a spatial scale known as the thermal wavelength such that $\lambda_{T}\propto\sqrt{\beta}$  (for example, in the case of an ideal gas, $\lambda_{T}=2h\sqrt{\frac{\beta}{2m}}$). Qualitatively, $\lambda_{T}$ is the minimum size of quantum wavepackets that describe the particles in a given system at a given temperature. Because of this relation between $\lambda_T$ and $\sqrt{\beta}$, we can focus on the dependence of $P_{\max}$ on $\sqrt{\beta}$.

Therefore, in Fig.~\ref{fig.Pmax}(a) we study the maximum probability $P_{\max}$ of localization for different values of $\sqrt{\beta}$ while fixing the size of the box $\Delta x$.  We localize in the region of size $\Delta x = 5$, and use the lattice sizes $L=10$, $20$, and $30$, with $N_p=3$ particles inside. We do this for both real and complex initial RPTSs.

We see that for cold systems (high $\beta$), the probability of localization is very small, in fact, $P_{\max}$ approaches zero. This is in accordance with the result of~\cite{deutsch2020probabilistic} which asserts that, in the limit of large $\sqrt{\beta}$ such that $\Delta x\ll \lambda$, $P_{\max}\propto (\frac{\Delta x}{\lambda_{T}})^{N_{p}d/2}$ where $d$ is the dimension of the lattice (in our case $d=1$). Intuitively, since $\lambda_{T}$ is the minimum size of quantum wavepackets, it makes sense that one can not localize the wavefunction in a subspace smaller than this length scale.

For hot systems (low $\beta$), the probability of localization $P_{\max}$ achieves high values. One notices that for small systems for e.g. 
$L=10$, the gap between $P_{\max}$ for the real and complex wave functions disappears. 
This is trivial, since in this case, the size of the subsystem of interest is becoming comparable to that of the full system, and therefore it is very easy to localize all particles in it.
For larger system sizes for e.g. $L=20,30$ 
all three conditions stated above are satisfied, and $P_{\max}$ approaches constant values of $\sim 1/2$ in the case of real RPTSs 
and $\sim  \pi^{2}/16$ in the case of complex RPTSs. 
The low $\beta$ regime is further explored in Fig.~\ref{fig.Pmax}(b).

To generate this graph, we used the same algorithm to maximize probability as the one used in~\cite{deutsch2020probabilistic}, and $\beta=0.01$. We start in $100$ different real and complex RPTSs, and for each one of them we perform the maximization procedure. We use three fermions and choose the small region to be $\Delta x=5$ sites. $P_{\max}$ is plotted as a function of $N/M^{2}$ for both cases of complex and real RPTSs.

As one can see, when the system is hot enough, it is possible to localize all the particles into the small region, and the probability that we find them there is at most $1/2$ and $\pi^{2}/16$ for real and complex RPTSs respectively. The presence of some fluctuations is expected since our model is a real system with non-random energy eigenvectors. This numerically confirms that the results of Deutsch et al.~\cite{deutsch2020probabilistic} also holds for a realistic quantum thermodynamic system such as ours, and we can apply this result in the next section.

\section{Role of localization in extreme values of entropies}\label{sec:Connection}
\begin{figure}[!t]
\includegraphics[width=1\linewidth]{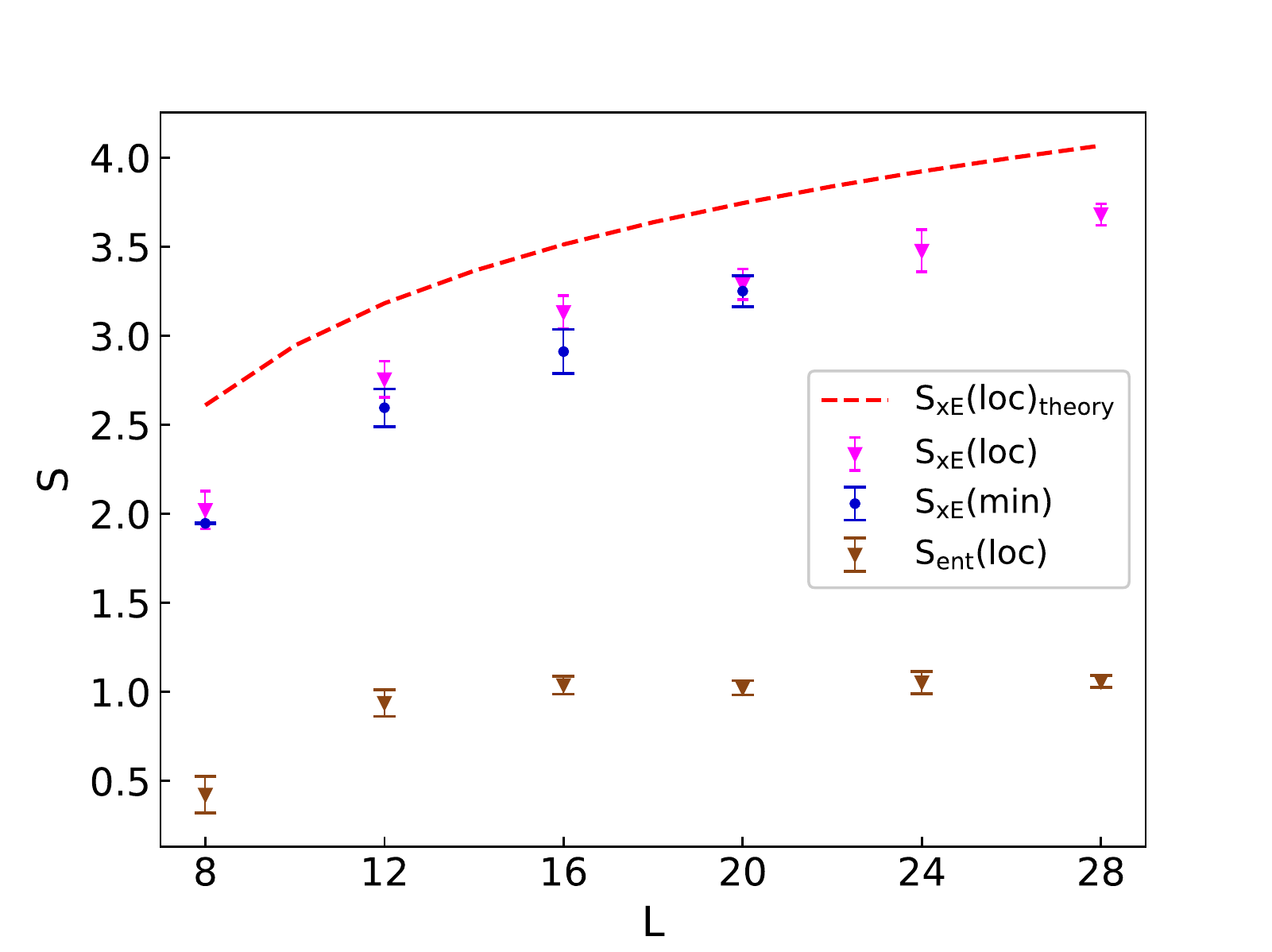}\\
\caption  {$N_p=2$ particles are localized in the first $4$ sites for varying system sizes $L$. $S_{\xE}$ and $S_{\ent}$ of such a state (i.e., when $P_{\max}$ is achieved) are computed and are respectively named $S_{\xE}(\loc)$ and $S_{\ent}(\loc)$. This is done for $6$ different complex RPTSs. What are illustrated here are the mean and standard deviation of these $6$ values. For comparison purposes, minimum values of $S_{\xE}$ achieved using optimization algorithm, $S_{\xE}(\min)$, are also plotted. Comparing these minima with corresponding $S_{\xE}(\loc)$ for each system size, one finds that these values are relatively very close to each other. This is an evidence of the key role of spatial localization in minimizing observational entropy. In contrast, entanglement entropy grows to a constant value.
This is related to the fact that entanglement entropy depends heavily on the distribution of particles for a given state. When the probability of localization is maximized for large system sizes, this distribution is fixed and independent of system size. We therefore expect entanglement entropy of such localized states to also reach a constant value  independent of system size. 
}
\label{fig.local_SxeSent}
\end{figure}

In figures~\ref{Hist_Sent} and~\ref{Hist_Sxe} we showed that minimizing entanglement and observational entropy leads to a substantial probability of localization in the larger and smaller regions respectively. 
In this section, we investigate what happens to entanglement entropy when one localizes particles into the small region as opposed to the bath, and the extent to which the spatial localization plays a role in minimizing the $S_{\xE}$.

We compute entropies of localized states, for $\Delta x =4$ and varying system sizes $L$. We consider $N_p=2$ particles in the system, and temperature is fixed at $\beta=0.01$, so that the three conditions from the previous section are satisfied. For each $L$, we start in $6$ initial complex RPTSs, and localize them into a physical region of fixed size $\Delta x=4$, by maximizing probability $P_{\max}$ for each initial state. We then compute the mean values and standard deviations of $S_{\ent}$ and $S_{\xE}$ of such localized states, and plot them in Fig.~\ref{fig.local_SxeSent}. The mean values of $P_{\max}$ (averaged over $6$ initial RPTSs) for system sizes of $L=[8,12,16,20,24,28]$ are $P_{\max}=[0.90,0.73,0.67,0.67,0.65,0.63]$.

$S_{\xE}(\loc)$ is very close to the minimum $S_{\xE}(\min)$ (discussed in detail in Fig.~\ref{fig.local_SxeSent}), showing that spatial localization is key in minimizing $S_{\xE}$. The theory predicts~\cite{deutsch2020probabilistic}
\[\label{Sxe_Sth}
\begin{split}
S_{\xE}(\loc)&=S_{\mathrm{th}}(L,N_p,\beta)\\
   &-P_{\max} N_p \ln \tfrac{L}{\Delta x}-(1-P_{\max})N_p\ln \tfrac{L}{L-\Delta x}\\
   &-P_{\max} \ln P_{\max}-(1-P_{\max})\ln(1-P_{\max}),
\end{split}
\]
for large $L$ (where $S_{\mathrm{th}}(L,N_p,\beta)\equiv S_{\mathrm{th}}(A+B)$), which is bounded below by
\[\label{eq:SxEgtHalfSth}
S_{\xE}(\loc) \geq (1-P_{\max}) S_{\mathrm{th}}(L,N_p,\beta),
\]
which shows that $S_{\xE}(\loc)$ cannot fall below a certain fraction of the total thermodynamic entropy of the system. Eq.~\eqref{Sxe_Sth} is plotted as a dashed line in Fig.~\ref{fig.local_SxeSent} and as expected from Eq.~\eqref{eq:SxEgtHalfSth}, the ratio $R=S_{\xE}(\loc)/S_{\mathrm{th}}(A+B)$ remains approximately constant for large $L$.

The fact that $S_{\xE}(\loc)$ and $S_{\xE}(\min)$ are almost the same and that $S_{\xE}(\min)$ is bounded by a fraction of thermodynamic entropy also explains why the minimum of $S_{\xE}$ in Fig.~\ref{Hist_Sxe} does not go to zero, and why $S_{\xE}(\min)$ in Fig.~\ref{fig.temperature_sxe} goes upwards for small $\beta$ (in the case of low $\beta$ $P_{\max}=\pi^2/16$).

\begin{figure}[t!]
\includegraphics[width=1\hsize]{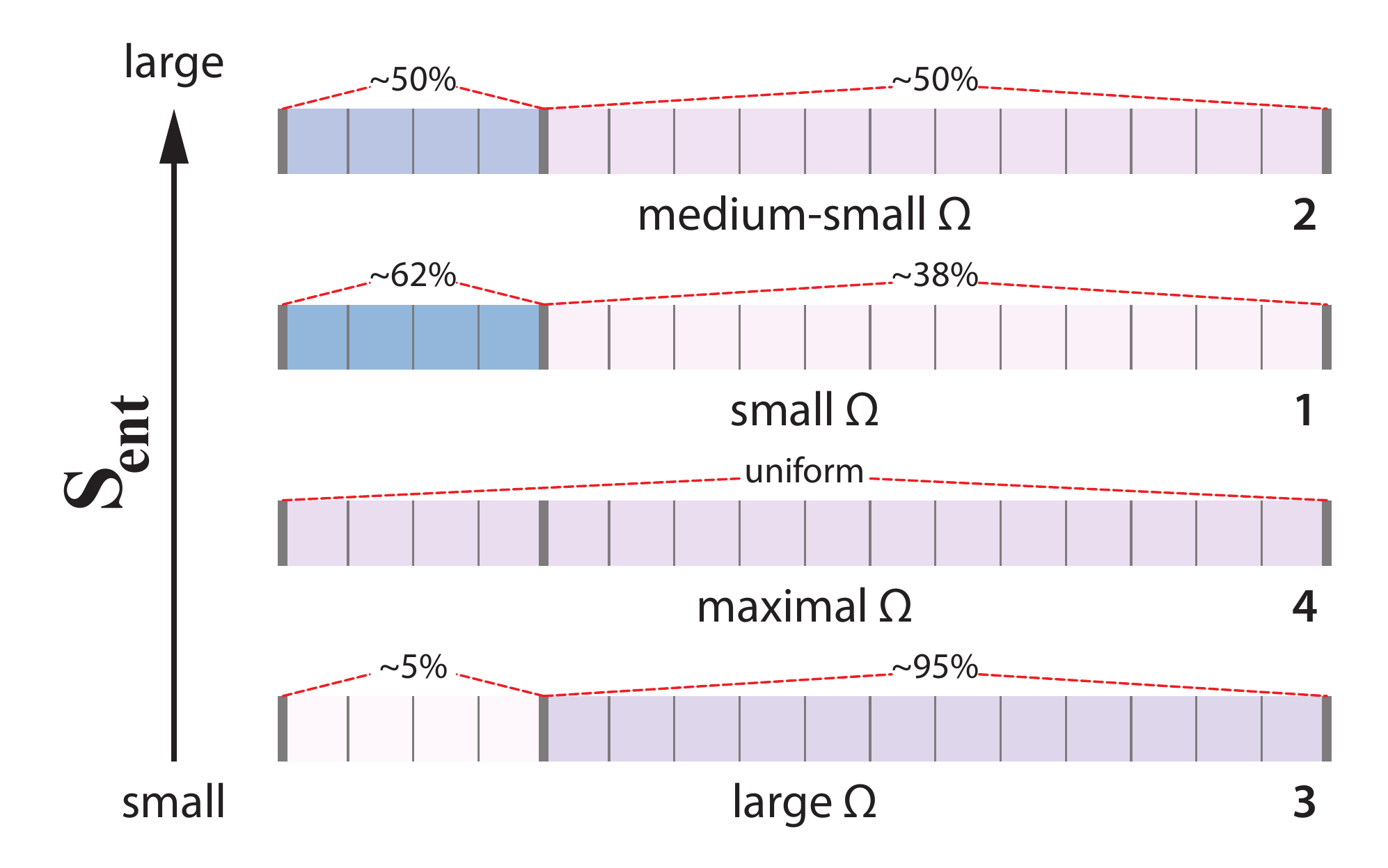}\\
~~~(a)
\includegraphics[width=1\hsize]{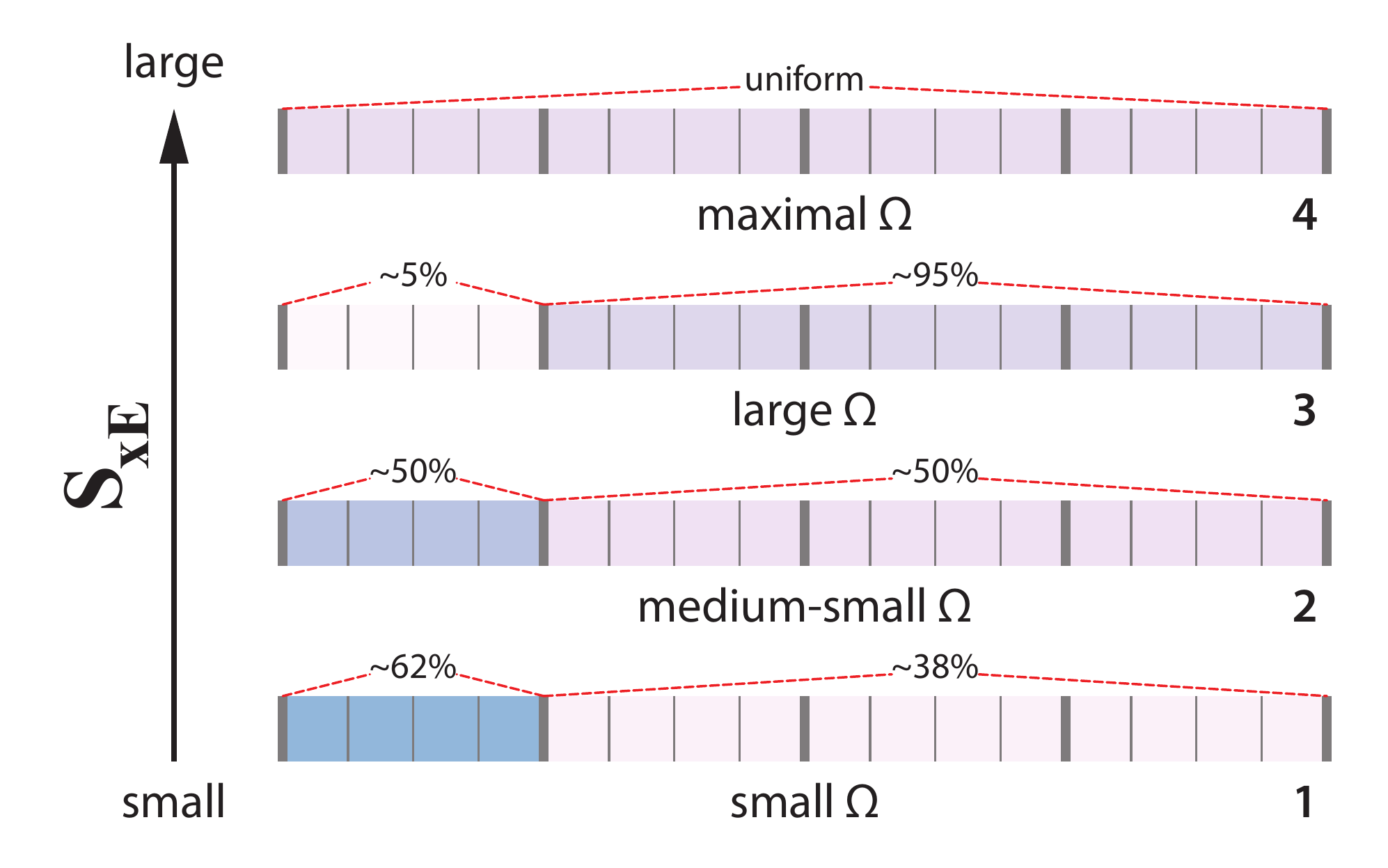}\\
~~~(b)
\caption{This illustration shows entropies for various types of macrostates described by distributions in their particle density, in order of  smaller to higher (a) entanglement entropy and smaller to higher (b) observational entropy. This is done as follows: we compute different types of microstates: state (1) of minimal $S_{xE}$, state (2) of maximal $S_{\ent}$, state (3) of minimal $S_{\ent}$, and equilibrium state (4) -- which is practically identical to the state of maximal $S_{xE}$.  We then plot their particle density and the percentages of the total number of particles in each subsystem. Each distribution of particle density defines a macrostate. The number of microstates that would lead to the same distribution of particle density defines the size of the macrostate, and is denoted $\Omega$. The Boltzmann entropy of a macrostate is then defined as $S_B=\ln \Omega$. In (a), the bottom lattice corresponds to the case of state (3) of minimal $S_{\ent}$: most particles are localized in the bath, and as a result the size of this macrostate is large compared to the other cases. The top lattice corresponds to the case with the state (2) of maximal $S_{\ent}$: 
In this case the average number of particles in sub-lattice $A$ follows condition~\eqref{eq:na_maxSent}.
The equal distribution of particles in the bath and subsystem shown here is a special case relevant to the numerical studies in this paper, but this is not in general the case. This configuration is different than that of the equilibrium state (4), in which case particles are distributed uniformly. In (b), larger macrostates correspond to larger observational entropy, showing correspondence with Boltzmann entropy $S_B=\ln \Omega$.
}
\label{fig.spec}
\end{figure}

\section{Discussion and conclusion}\label{sec:conclusion}
In this paper, we have demonstrated the significantly different behavior of extreme values of entanglement and observational entropy by illustrating the behaviour of these entropies in and out of equilibrium and at different temperatures in an isolated quantum system. In particular we show that observational entropy behaves similarly to Boltzmann entropy, even in its extreme cases, i.e., far from equilibrium, while entanglement entropy does not. Indeed these entropies pertain to rather different physical features: observational entropy has the ability to quantify localization of particles and energy, while entanglement entropy measures non-local correlations, for systems both in and out of equilibrium. 

With regards to extreme values, we found that starting from a random pure thermal state, $S_{\ent}$ can reach values very close to zero during the course of a unitary evolution, whereas there exists a non-zero lower bound for $S_{\xE}$. We showed how these minimal values of the two entropies are achieved through localization in the larger and smaller region for  $S_{\ent}$ and $S_{\xE}$, respectively.

We found that in the high temperature limit, the maximum entanglement entropy between a smaller subsystem and the rest of the system, becomes very large in comparison with typical values, and this ratio grows with system size. This is because the typical value is the thermodynamic entropy for the smaller subsystem, which will tend to zero as total system size increases (at fixed particle number and subsystem size) but the maximum $S_{ent}$ goes to a constant independent of total system size. On the other hand, the maximal observational entropy stays close to its typical value. The latter is qualitatively similar to classical Boltzmann entropy: the average -- which is close to the most likely state -- should be assigned a very high entropy.

The particle distribution given a state with maximum entanglement or observational entropy is also markedly different: in the former case, the particles distribute themselves throughout the lattice such that the average number of particles in the subsystem 
follows condition~\eqref{eq:na_maxSent} in pursuit of maximizing correlations between the two subsystems, Whereas in the latter case, particles tend to distribute themselves uniformly, similar to what happens at thermal equilibrium.

These findings are illustrated in Fig.~\ref{fig.spec}.  In particular, Fig.~\ref{fig.spec} (a) and (b) 
show entropies for various types of macrostates, described by their particle density, in order of smaller to higher entanglement and observational entropy, respectively. 
From the Boltzmann point of view, the size of the macrostate is determined by the number of microstates corresponding to the same macroscopic appearance: in this figure, size of the macrostate $\Omega$ is the number of orthogonal quantum states that give the same distribution of particle density.

One notices that higher entanglement entropy does not necessarily mean that the macrostate is larger -- the size of the macrostate appears to be rather unrelated to the amount of entanglement entropy. Specifically, it would be more likely to observe a state with minimal entanglement entropy as compared to the maximal entanglement entropy (as the former has a larger macrostate). The size of the macrostate and the entropy of the state match for the case of observational entropy, showing that (quantum) observational entropy matches well with the classical conception of Boltzmann entropy.

It should be noted that in this paper we focused on bipartite entanglement entropy, since it is very often used in literature. One could argue that multipartite entanglement entropy, defined as the sum of local von Neumann entropies, could behave similarly to $S_{\xE}$ and be more Boltzmann-like, meaning that the larger macrostates have associated higher values of entropy. \footnote{This property would be however dependent upon having equally-sized regions, and the equivalence would break even when the size of a single region is different from others.}

Because of its close relation to Boltzmann entropy, observational entropy could accompany the entanglement entropy to better understand the concept of thermalization in isolated quantum systems, and to illuminate the behavior of out-of-equilibrium states which lie at the heart of statistical mechanics. This entropy is also rather new in the field of quantum thermodynamics and hence further work on this particular entropy is of interest. 

Experimentally, for example, observational entropy could be measured without the need to access the density matrix, and be useful in quantifying how thermalized a given state is. The extreme values of this entropy could possibly be probed as well, for extremely small systems such that the time it takes to reach these values is reasonably small and within reach in laboratories. 

Finally, on a cosmological level,
discussions of entropy and the arrow of time (e.g.~\cite{davies1977physics,aguirre2003inflation,albrecht2004can,penrose2004road,carroll2004spontaneous,aguirre2012out}) require a definition that applies to a truly closed system (like the Universe), out of equilibrium, and potentially for indefinitely long timescales over which large entropy fluctuations might occur. These discussions often employ an ``informal" definition of entropy that in practice mixes different notions. Observational entropy applies in this context and is rigorously defined, and therefore may be very useful in these discussions. (The primarily remaining obstacle being a lack of understanding of the state-space of gravity and spacetime.) Could this definition of entropy, for example, tell us something new about the arrow of time in isolated quantum systems, and about how to understand extreme entropy fluctuations in the context of the arrow of time?

Extending this work to other types of observational entropies that are in accordance with thermodynamic entropy (such as ``Factorized Observational Entropy''~\cite{safranek2019long}) would be of interest as well. This could give us a broader understanding of this 
definition of entropy and non-equilibrium many-body quantum systems.

\acknowledgements{This research was supported by the Foundational Questions Institute (FQXi.org) of which AA is Associate Director, and the Faggin Presidential Chair Fund.}

\appendix
\section{Short introduction to observational entropy}\label{app:oe}

Observational entropy is a generalization of Boltzmann entropy into quantum systems. It was mentioned but not developed by von Neumann in 1929~\cite{von2010proof,von1955mathematical} then more recently rediscovered, developed, generalized to include multiple (even non-commuting) coarse-grainings, and connected to thermodynamics in~\cite{safranek2019letter,safranek2019long}, using the following construction:

Let us assume that the Hilbert space can be decomposed into a direct sum of orthogonal subspaces $\HS=\bigoplus_i\HS_i$, where each subspace corresponds to a macrostate specifying a single macroscopic property of the system (such as energy or number of particles). Defining projector $\P_i$ as the projector onto a subspace $\HS_i$, the set $\C=\{\P_i\}$ forms a trace-preserving set of orthogonal projectors, denoted a coarse-graining. The probability that a quantum state $\R$ is in a given macrostate can be calculated as $p_i=\tr[\P_i\R]$. Equivalently, we can say that this is the probability that a system described by a quantum state $\R$ will be found having a value $i$ of a macroscopic property defined by the coarse-graining, when performing a coarse-grained measurement on it in the basis given by the coarse-graining.

Assuming that an observer cannot distinguish between different microstates $k$ within the same macrostate $i$ with his/her macroscopic measurement, he/she associates the same probability $p_i^{(k)}=p_i/\dim(\HS_i)=p_i/\tr[\P_i]$ to every microstate (given by a pure quantum state) in the macrostate. Given this inability to distinguish between different microstates within the same macrostate, we consider the Shannon entropy of the probabilities $p_i^{(k)}$,
\[
S_{O(\C)}\equiv-\sum_{i,k} p_i^{(k)}\ln p_i^{(k)}=-\sum_{i}p_i\ln \frac{p_i}{\tr[\P_i]},
\]
This defines observational entropy with a single coarse-graining.

A generalization of this quantity for multiple coarse-grainings that allows many of its properties to be retained is
\[
S_{O(\C_1,\dots,\C_n)}\equiv-\sum_{\bi}p_\bi\ln \frac{p_\bi}{V_\bi},
\]
where multi-index $\bi=(i_1,\dots,i_n)$ denotes a set of macroscopic properties, $p_\bi$ is the probability of these properties being measured (in the given order), and $V_\bi=\tr[\P_{i_n}\cdots\P_{i_1}\cdots\P_{i_n}]$ denotes a joint Hilbert space volume of all systems that have properties $\bi=(i_1,\dots,i_n)$ measured in this order. Equivalently, we can call $V_\bi$ the volume of multi-macrostate $\bi$.

An important property of $S_O$ is that it depends on the order of coarse-grainings, and that for any ordered set of coarse-grainings $(\C_1,\dots,\C_n)$ and any density matrix $\R$,
\begin{align} 
S_{V\!N}(\R)\leq S_{O(\C_1,\dots,\C_n)}(\R)\leq \ln \mathrm{dim}\HS,\\ 
S_{O(\C_1,\dots,\C_n)}(\R)\leq 
\label{eq:nonincreasing} S_{O(\C_1,\dots,\C_{n-1})}(\R).  
\end{align}
In words, this means that observational entropy is lower bounded by the von Neumann entropy, which can be interpreted as an inherent uncertainty in a quantum system, upper bounded by the maximal uncertainty in the system, and that with each added coarse-graining, the observational entropy does not increase. These properties show that observational entropy can be interpreted as an observers' uncertainty about the system, given that all that he/she can do is to perform a set of macroscopic measurements on this system.

Despite the intuitive interpretation of this general quantity, its physical meaning depends upon the coarse-graining.  Several pertinent examples we have identified are as follows.

First is a coarse-graining in energy, $\C_E=\{\P_E\}_E$, where $\P_E$ is a projector onto a subspace associated with eigenvalue $E$ of the Hamiltonian $\hat{H}$ (for non-degenerate Hamiltonians, $\P_E=\ket{E}\bra{E}$ is a projector onto a single energy eigenstate). Observational entropy $S_E\equiv S_{O(\C_E)}$ then gives the equilibrium value of the thermodynamic entropy. For example, for a microcanonical state $\R_{\mathrm{micro}}=\frac{1}{N(E)}\sum_{E\leq\tilde{E}<E+\Delta E}\ket{\tilde{E}}\bra{\tilde{E}}$ it gives the microcanonical entropy, 
\[
S_E(\R_{\mathrm{micro}})=\ln(N(E))=\ln(\rho(E)\Delta E)=S_{\mathrm{micro}}(E),
\]
where $\rho(E)$ denotes the energy density of states, while for the canonical state $\R_{th}=\frac{1}{Z}\exp(-\beta \hat{H})$ it gives the canonical entropy,
\[
S_E(\R_{th})=\ln Z+\beta\mean{E}=S_{V\!N}(\R_{th})=S_{th}.
\]

A second example, leading to time-dependence while still pertaining to thermodynamics, is the coarse-graining in local Hamiltonians, $\C_{\hat{H}_1}\otimes\cdots\otimes\C_{\hat{H}_m}=\{\P_{E_1,\dots,E_m}\}$, where $\P_{E_1,\dots,E_m}=\P_{E_1}\otimes\cdots\otimes \P_{E_m}$, $\hat{H}_j\P_{E_j}=E_j\P_{E_j}$, are projectors onto local energy eigenstates. The resulting entropy $S_F\equiv S_{O(\C_{\hat{H}_1}\otimes\cdots\otimes\C_{\hat{H}_m})}$ then measures how close to equilibrium these local regions are. 

A third and closely-related example is $S_{\xE}$, Eq.~\eqref{SxE}, which is the main focus of this paper. This observational entropy, coarse-grained first in local particle numbers, and then in total energy, can be interpreted as entropy that an observer would associate to a system where $m$ partitions are allowed to exchange energy but not particles, in the long-time limit. In our paper the particles do indeed exchange between the partitions, which is the reason why this quantity is time-dependent. $S_{F}$ is upper bounded by thermodynamic entropy (either canonical or microcanonical, depending on the spread in energies), and converges to it in the long-time limit in any non-integrable closed quantum system, up to small corrections (see Ref.~\cite{safranek2019long}). $S_{\xE}$ is upper bounded by thermodynamic entropy and converges to it in the long-time limit only in the high-temperature limit (as can be seen in Fig.~\ref{fig.temperature_sxe}). A few other related entropies can be defined by considering a finite resolution in energy; these come with some subtleties, and resolve some slight issues in the behavior of the current definitions.\footnote{Note that switching the order of coarse-graining, leads to an entropy $S_{\mathrm{Ex}}$ that is always lower than the thermodynamic entropy in both high and low-temperature limit as per Eq.~\eqref{eq:nonincreasing}, $S_{\mathrm{Ex}}\leq S_E(\R)\leq S_{th}$, but comes with its own set of problems (for example, it associates zero entropy to any energy eigenstate.) This can, however, be  resolved by introducing a suitable resolution in measuring energy $\Delta E$ (not too low and not too high), which then leads to an entropy $S_{\mathrm{Ex}}^{(\Delta E)}$ that associates microcanonical entropy to energy eigenstates, and an entropy that converges to $S_E(\R)\approx S_{th}$ in the long-time limit for states that are superpositions of many energy eigenstates. The fact that a suitable $\Delta E$ makes this entropy useful has been only discovered recently by us and only briefly touched upon in~\cite{safranek2019classical}, and is not discussed in the original papers~\cite{safranek2019letter,safranek2019long}, which in Appendix H of~\cite{safranek2019long} claimed that $S_{\mathrm{Ex}}^{(\Delta E)}$ is not a good definition of entropy (while it indeed is, as we have just explained). A similar problem happens with $S_F$, which associates zero entropy to states consisting of local energy eigenstates, $S_F(\ket{E_1}\otimes\cdots\otimes\ket{E_m})=0$. Again, this can be resolved by introducing small but non-zero resolution in measuring local energies $\Delta E$, which then defines an entropy that leads to the sum of local microcanonical entropies, $S_F^{(\Delta E)}(\ket{E_1}\otimes\cdots\otimes\ket{E_m})=\sum_{i=1}^m S_{\mathrm{micro}}(E_i)$.}

We chose to study $S_{\xE}$ in this paper, because it is significantly easier to calculate, and since in most cases (especially in the high-temperature limit), all $S_{\xE}$, $S_F$, $S_{\mathrm{Ex}}^{(\Delta E)}$ behave quite similarly.

\bibliographystyle{apsrev}
\bibliography{bibl7}

\begin{thebibliography}{71}
\expandafter\ifx\csname natexlab\endcsname\relax\def\natexlab#1{#1}\fi
\expandafter\ifx\csname bibnamefont\endcsname\relax
  \def\bibnamefont#1{#1}\fi
\expandafter\ifx\csname bibfnamefont\endcsname\relax
  \def\bibfnamefont#1{#1}\fi
\expandafter\ifx\csname citenamefont\endcsname\relax
  \def\citenamefont#1{#1}\fi
\expandafter\ifx\csname url\endcsname\relax
  \def\url#1{\texttt{#1}}\fi
\expandafter\ifx\csname urlprefix\endcsname\relax\def\urlprefix{URL }\fi
\providecommand{\bibinfo}[2]{#2}
\providecommand{\eprint}[2][]{\url{#2}}

\bibitem[{\citenamefont{Page}(1993)}]{page1993average}
\bibinfo{author}{\bibfnamefont{D.~N.} \bibnamefont{Page}},
  \bibinfo{journal}{\href{https://journals.aps.org/prl/abstract/10.1103/PhysRevLett.71.1291}{Phys.
  Rev. Lett.}} \textbf{\bibinfo{volume}{71}}, \bibinfo{pages}{1291}
  (\bibinfo{year}{1993}).

\bibitem[{\citenamefont{Grasselli}(2014)}]{grassellientanglement}
\bibinfo{author}{\bibfnamefont{F.}~\bibnamefont{Grasselli}}, Master's thesis,
  \bibinfo{school}{Universit\`a degli Studi di Perugia} (\bibinfo{year}{2014}).

\bibitem[{\citenamefont{Vidmar et~al.}(2017)\citenamefont{Vidmar, Hackl,
  Bianchi, and Rigol}}]{vidmar2017entanglementquadratic}
\bibinfo{author}{\bibfnamefont{L.}~\bibnamefont{Vidmar}},
  \bibinfo{author}{\bibfnamefont{L.}~\bibnamefont{Hackl}},
  \bibinfo{author}{\bibfnamefont{E.}~\bibnamefont{Bianchi}}, \bibnamefont{and}
  \bibinfo{author}{\bibfnamefont{M.}~\bibnamefont{Rigol}},
  \bibinfo{journal}{\href{https://journals.aps.org/prl/abstract/10.1103/PhysRevLett.119.020601}{Phys.
  Rev. Lett.}} \textbf{\bibinfo{volume}{119}}, \bibinfo{pages}{020601}
  (\bibinfo{year}{2017}).

\bibitem[{\citenamefont{Vidmar and Rigol}(2017)}]{vidmar2017entanglement}
\bibinfo{author}{\bibfnamefont{L.}~\bibnamefont{Vidmar}} \bibnamefont{and}
  \bibinfo{author}{\bibfnamefont{M.}~\bibnamefont{Rigol}},
  \bibinfo{journal}{\href{https://journals.aps.org/prl/abstract/10.1103/PhysRevLett.119.220603}{Phys.
  Rev. Lett.}} \textbf{\bibinfo{volume}{119}}, \bibinfo{pages}{220603}
  (\bibinfo{year}{2017}).

\bibitem[{\citenamefont{Deutsch et~al.}(2013)\citenamefont{Deutsch, Li, and
  Sharma}}]{deutsch2013microscopic}
\bibinfo{author}{\bibfnamefont{J.~M.} \bibnamefont{Deutsch}},
  \bibinfo{author}{\bibfnamefont{H.}~\bibnamefont{Li}}, \bibnamefont{and}
  \bibinfo{author}{\bibfnamefont{A.}~\bibnamefont{Sharma}},
  \bibinfo{journal}{\href{https://journals.aps.org/pre/abstract/10.1103/PhysRevE.87.042135}{Phys.
  Rev. E}} \textbf{\bibinfo{volume}{87}}, \bibinfo{pages}{042135}
  (\bibinfo{year}{2013}).

\bibitem[{\citenamefont{Zhang et~al.}(2015)\citenamefont{Zhang, Kim, and
  Huse}}]{zhang2015thermalization}
\bibinfo{author}{\bibfnamefont{L.}~\bibnamefont{Zhang}},
  \bibinfo{author}{\bibfnamefont{H.}~\bibnamefont{Kim}}, \bibnamefont{and}
  \bibinfo{author}{\bibfnamefont{D.~A.} \bibnamefont{Huse}},
  \bibinfo{journal}{\href{https://journals.aps.org/pre/abstract/10.1103/PhysRevE.91.062128}{Phys.
  Rev. E}} \textbf{\bibinfo{volume}{91}}, \bibinfo{pages}{062128}
  (\bibinfo{year}{2015}).

\bibitem[{\citenamefont{Kaufman et~al.}(2016)\citenamefont{Kaufman, Tai, Lukin,
  Rispoli, Schittko, Preiss, and Greiner}}]{kaufman2016quantum}
\bibinfo{author}{\bibfnamefont{A.~M.} \bibnamefont{Kaufman}},
  \bibinfo{author}{\bibfnamefont{M.~E.} \bibnamefont{Tai}},
  \bibinfo{author}{\bibfnamefont{A.}~\bibnamefont{Lukin}},
  \bibinfo{author}{\bibfnamefont{M.}~\bibnamefont{Rispoli}},
  \bibinfo{author}{\bibfnamefont{R.}~\bibnamefont{Schittko}},
  \bibinfo{author}{\bibfnamefont{P.~M.} \bibnamefont{Preiss}},
  \bibnamefont{and} \bibinfo{author}{\bibfnamefont{M.}~\bibnamefont{Greiner}},
  \bibinfo{journal}{\href{https://science.sciencemag.org/content/353/6301/794}{Science}}
  \textbf{\bibinfo{volume}{353}}, \bibinfo{pages}{794} (\bibinfo{year}{2016}).

\bibitem[{\citenamefont{Osborne and Nielsen}(2002)}]{osborne2002entanglement}
\bibinfo{author}{\bibfnamefont{T.~J.} \bibnamefont{Osborne}} \bibnamefont{and}
  \bibinfo{author}{\bibfnamefont{M.~A.} \bibnamefont{Nielsen}},
  \bibinfo{journal}{\href{https://journals.aps.org/pra/abstract/10.1103/PhysRevA.66.032110}{Phys.
  Rev. A}} \textbf{\bibinfo{volume}{66}}, \bibinfo{pages}{032110}
  (\bibinfo{year}{2002}).

\bibitem[{\citenamefont{Latorre et~al.}(2003)\citenamefont{Latorre, Rico, and
  Vidal}}]{latorre2003ground}
\bibinfo{author}{\bibfnamefont{J.~I.} \bibnamefont{Latorre}},
  \bibinfo{author}{\bibfnamefont{E.}~\bibnamefont{Rico}}, \bibnamefont{and}
  \bibinfo{author}{\bibfnamefont{G.}~\bibnamefont{Vidal}},
  \bibinfo{journal}{\href{https://dl.acm.org/doi/10.5555/2011572.2011576}{Quant.
  Inf. Comput.}} \textbf{\bibinfo{volume}{4}}, \bibinfo{pages}{48}
  (\bibinfo{year}{2003}).

\bibitem[{\citenamefont{Vidal et~al.}(2003)\citenamefont{Vidal, Latorre, Rico,
  and Kitaev}}]{vidal2003entanglement}
\bibinfo{author}{\bibfnamefont{G.}~\bibnamefont{Vidal}},
  \bibinfo{author}{\bibfnamefont{J.~I.} \bibnamefont{Latorre}},
  \bibinfo{author}{\bibfnamefont{E.}~\bibnamefont{Rico}}, \bibnamefont{and}
  \bibinfo{author}{\bibfnamefont{A.}~\bibnamefont{Kitaev}},
  \bibinfo{journal}{\href{https://journals.aps.org/prl/abstract/10.1103/PhysRevLett.90.227902}{Phys.
  Rev. Lett.}} \textbf{\bibinfo{volume}{90}}, \bibinfo{pages}{227902}
  (\bibinfo{year}{2003}).

\bibitem[{\citenamefont{Emparan}(2006)}]{emparan2006black}
\bibinfo{author}{\bibfnamefont{R.}~\bibnamefont{Emparan}},
  \bibinfo{journal}{\href{https://iopscience.iop.org/article/10.1088/1126-6708/2006/06/012/meta}{J.
  High Energy Phys.}} \textbf{\bibinfo{volume}{2006}}, \bibinfo{pages}{012}
  (\bibinfo{year}{2006}).

\bibitem[{\citenamefont{Nishioka}(2018)}]{nishioka2018entanglement}
\bibinfo{author}{\bibfnamefont{T.}~\bibnamefont{Nishioka}},
  \bibinfo{journal}{\href{https://journals.aps.org/rmp/abstract/10.1103/RevModPhys.90.035007}{Rev.
  Mod. Phys.}} \textbf{\bibinfo{volume}{90}}, \bibinfo{pages}{035007}
  (\bibinfo{year}{2018}).

\bibitem[{\citenamefont{Mukohyama}(1998)}]{mukohyama1998comments}
\bibinfo{author}{\bibfnamefont{S.}~\bibnamefont{Mukohyama}},
  \bibinfo{journal}{\href{https://journals.aps.org/prd/abstract/10.1103/PhysRevD.97.066025}{Phys.
  Rev. D}} \textbf{\bibinfo{volume}{58}}, \bibinfo{pages}{104023}
  (\bibinfo{year}{1998}).

\bibitem[{\citenamefont{Unanyan and
  Fleischhauer}(2005)}]{unanyan2005entanglement}
\bibinfo{author}{\bibfnamefont{R.~G.} \bibnamefont{Unanyan}} \bibnamefont{and}
  \bibinfo{author}{\bibfnamefont{M.}~\bibnamefont{Fleischhauer}},
  \bibinfo{journal}{\href{https://journals.aps.org/prl/abstract/10.1103/PhysRevLett.95.260604}{Phys.
  Rev. Lett.}} \textbf{\bibinfo{volume}{95}}, \bibinfo{pages}{260604}
  (\bibinfo{year}{2005}).

\bibitem[{\citenamefont{Cho}(2018)}]{cho2018realistic}
\bibinfo{author}{\bibfnamefont{J.}~\bibnamefont{Cho}},
  \bibinfo{journal}{\href{https://journals.aps.org/prx/abstract/10.1103/PhysRevX.8.031009}{
  Phys. Rev. X}} \textbf{\bibinfo{volume}{8}}, \bibinfo{pages}{031009}
  (\bibinfo{year}{2018}).

\bibitem[{\citenamefont{\ifmmode~\check{S}\else \v{S}\fi{}afr\'anek
  et~al.}(2019{\natexlab{a}})\citenamefont{\ifmmode~\check{S}\else
  \v{S}\fi{}afr\'anek, Deutsch, and Aguirre}}]{safranek2019letter}
\bibinfo{author}{\bibfnamefont{D.}~\bibnamefont{\ifmmode~\check{S}\else
  \v{S}\fi{}afr\'anek}}, \bibinfo{author}{\bibfnamefont{J.~M.}
  \bibnamefont{Deutsch}}, \bibnamefont{and}
  \bibinfo{author}{\bibfnamefont{A.}~\bibnamefont{Aguirre}},
  \bibinfo{journal}{\href{https://link.aps.org/doi/10.1103/PhysRevA.99.010101}{Phys.
  Rev. A}} \textbf{\bibinfo{volume}{99}}, \bibinfo{pages}{010101(R)}
  (\bibinfo{year}{2019}{\natexlab{a}}).

\bibitem[{\citenamefont{\ifmmode~\check{S}\else \v{S}\fi{}afr\'anek
  et~al.}(2019{\natexlab{b}})\citenamefont{\ifmmode~\check{S}\else
  \v{S}\fi{}afr\'anek, Deutsch, and Aguirre}}]{safranek2019long}
\bibinfo{author}{\bibfnamefont{D.}~\bibnamefont{\ifmmode~\check{S}\else
  \v{S}\fi{}afr\'anek}}, \bibinfo{author}{\bibfnamefont{J.~M.}
  \bibnamefont{Deutsch}}, \bibnamefont{and}
  \bibinfo{author}{\bibfnamefont{A.}~\bibnamefont{Aguirre}},
  \bibinfo{journal}{\href{https://link.aps.org/doi/10.1103/PhysRevA.99.012103}{Phys.
  Rev. A}} \textbf{\bibinfo{volume}{99}}, \bibinfo{pages}{012103}
  (\bibinfo{year}{2019}{\natexlab{b}}).

\bibitem[{\citenamefont{{\v{S}}afr{\'a}nek
  et~al.}(2019)\citenamefont{{\v{S}}afr{\'a}nek, Aguirre, and
  Deutsch}}]{safranek2019classical}
\bibinfo{author}{\bibfnamefont{D.}~\bibnamefont{{\v{S}}afr{\'a}nek}},
  \bibinfo{author}{\bibfnamefont{A.}~\bibnamefont{Aguirre}}, \bibnamefont{and}
  \bibinfo{author}{\bibfnamefont{J.~M.} \bibnamefont{Deutsch}},
  \bibinfo{journal}{\href{https://arxiv.org/abs/1905.03841}{arXiv:1905.03841}
  [cond-mat.stat-mech]}  (\bibinfo{year}{2019}).

\bibitem[{\citenamefont{von Neumann}(2010)}]{von2010proof}
\bibinfo{author}{\bibfnamefont{J.}~\bibnamefont{von Neumann}},
  \bibinfo{journal}{\href{http://dx.doi.org/10.1140/epjh/e2010-00008-5}{Eur.
  Phys. J. H}} \textbf{\bibinfo{volume}{35}}, \bibinfo{pages}{201}
  (\bibinfo{year}{2010}).

\bibitem[{\citenamefont{von Neumann}(1955)}]{von1955mathematical}
\bibinfo{author}{\bibfnamefont{J.}~\bibnamefont{von Neumann}},
  \emph{\bibinfo{title}{Mathematical foundations of quantum mechanics}}
  (\bibinfo{publisher}{\href{http://press.princeton.edu/titles/2113.html}{Princeton
  university press}}, \bibinfo{address}{Princeton}, \bibinfo{year}{1955}), pp.
  \bibinfo{pages}{410--416}.

\bibitem[{\citenamefont{Wehrl}(1978)}]{wehrl1978general}
\bibinfo{author}{\bibfnamefont{A.}~\bibnamefont{Wehrl}},
  \bibinfo{journal}{\href{https://journals.aps.org/rmp/abstract/10.1103/RevModPhys.50.221}{Rev.
  Mod. Phys}} \textbf{\bibinfo{volume}{50}}, \bibinfo{pages}{221}
  (\bibinfo{year}{1978}).

\bibitem[{\citenamefont{Lent}(2019)}]{lent2019quantum}
\bibinfo{author}{\bibfnamefont{C.~S.} \bibnamefont{Lent}},
  \bibinfo{journal}{\href{https://journals.aps.org/pre/abstract/10.1103/PhysRevE.100.012101}{Phys.
  Rev. E}} \textbf{\bibinfo{volume}{100}}, \bibinfo{pages}{012101}
  (\bibinfo{year}{2019}).

\bibitem[{\citenamefont{Goldstein et~al.}(2019)\citenamefont{Goldstein,
  Lebowitz, Tumulka, and Zanghi}}]{goldstein2019gibbs}
\bibinfo{author}{\bibfnamefont{S.}~\bibnamefont{Goldstein}},
  \bibinfo{author}{\bibfnamefont{J.~L.} \bibnamefont{Lebowitz}},
  \bibinfo{author}{\bibnamefont{Tumulka}}, \bibnamefont{and}
  \bibinfo{author}{\bibfnamefont{N.}~\bibnamefont{Zanghi}},
  \bibinfo{journal}{\href{https://arxiv.org/abs/1903.11870}{arXiv:1903.11870}
  [cond-mat.stat-mech]}  (\bibinfo{year}{2019}).

\bibitem[{\citenamefont{Gemmer and Steinigeweg}(2014)}]{gemmer2014entropy}
\bibinfo{author}{\bibfnamefont{J.}~\bibnamefont{Gemmer}} \bibnamefont{and}
  \bibinfo{author}{\bibfnamefont{R.}~\bibnamefont{Steinigeweg}},
  \bibinfo{journal}{\href{https://journals.aps.org/pre/abstract/10.1103/PhysRevE.89.042113}{Phys.
  Rev. E}} \textbf{\bibinfo{volume}{89}}, \bibinfo{pages}{042113}
  (\bibinfo{year}{2014}).

\bibitem[{\citenamefont{Strasberg}(2019)}]{strasberg2019entropy}
\bibinfo{author}{\bibfnamefont{P.}~\bibnamefont{Strasberg}},
  \bibinfo{journal}{\href{https://arxiv.org/abs/1906.09933}{arXiv:1906.09933}
  [cond-mat.stat-mech]}  (\bibinfo{year}{2019}).

\bibitem[{\citenamefont{Boltzmann}(2003)}]{boltzmann2003further}
\bibinfo{author}{\bibfnamefont{L.}~\bibnamefont{Boltzmann}}, in
  \emph{\bibinfo{booktitle}{The kinetic theory of gases: an anthology of
  classic papers with historical commentary}} (\bibinfo{publisher}{World
  Scientific}, \bibinfo{address}{Hackensack}, \bibinfo{year}{2003}), pp.
  \bibinfo{pages}{262--349}.

\bibitem[{\citenamefont{Dunkel and Hilbert}(2014)}]{dunkel2014consistent}
\bibinfo{author}{\bibfnamefont{J.}~\bibnamefont{Dunkel}} \bibnamefont{and}
  \bibinfo{author}{\bibfnamefont{S.}~\bibnamefont{Hilbert}},
  \bibinfo{journal}{\href{https://math.mit.edu/~dunkel/Papers/2014DuHi_NatPhys.pdf}{Nat.
  Phys.}} \textbf{\bibinfo{volume}{10}}, \bibinfo{pages}{67}
  (\bibinfo{year}{2014}).

\bibitem[{\citenamefont{Hilbert et~al.}(2014)\citenamefont{Hilbert, H{\"a}nggi,
  and Dunkel}}]{hilbert2014thermodynamic}
\bibinfo{author}{\bibfnamefont{S.}~\bibnamefont{Hilbert}},
  \bibinfo{author}{\bibfnamefont{P.}~\bibnamefont{H{\"a}nggi}},
  \bibnamefont{and} \bibinfo{author}{\bibfnamefont{J.}~\bibnamefont{Dunkel}},
  \bibinfo{journal}{\href{https://journals.aps.org/pre/abstract/10.1103/PhysRevE.90.062116}{Phys.
  Rev. E}} \textbf{\bibinfo{volume}{90}}, \bibinfo{pages}{062116}
  (\bibinfo{year}{2014}).

\bibitem[{\citenamefont{Schneider et~al.}(2014)\citenamefont{Schneider, Mandt,
  Rapp, Braun, Weimer, Bloch, and Rosch}}]{schneider2014comment}
\bibinfo{author}{\bibfnamefont{U.}~\bibnamefont{Schneider}},
  \bibinfo{author}{\bibfnamefont{S.}~\bibnamefont{Mandt}},
  \bibinfo{author}{\bibfnamefont{A.}~\bibnamefont{Rapp}},
  \bibinfo{author}{\bibfnamefont{S.}~\bibnamefont{Braun}},
  \bibinfo{author}{\bibfnamefont{H.}~\bibnamefont{Weimer}},
  \bibinfo{author}{\bibfnamefont{I.}~\bibnamefont{Bloch}}, \bibnamefont{and}
  \bibinfo{author}{\bibfnamefont{A.}~\bibnamefont{Rosch}},
  \bibinfo{journal}{\href{https://arxiv.org/abs/1407.4127}{arXiv:1407.4127}
  [cond-mat.quant-gas]}  (\bibinfo{year}{2014}).

\bibitem[{\citenamefont{Buonsante et~al.}(2016)\citenamefont{Buonsante,
  Franzosi, and Smerzi}}]{buonsante2016dispute}
\bibinfo{author}{\bibfnamefont{P.}~\bibnamefont{Buonsante}},
  \bibinfo{author}{\bibfnamefont{R.}~\bibnamefont{Franzosi}}, \bibnamefont{and}
  \bibinfo{author}{\bibfnamefont{A.}~\bibnamefont{Smerzi}},
  \bibinfo{journal}{\href{https://www.sciencedirect.com/science/article/pii/S0003491616302342}{Ann.
  Physics}} \textbf{\bibinfo{volume}{375}}, \bibinfo{pages}{414}
  (\bibinfo{year}{2016}).

\bibitem[{\citenamefont{Goldstein and Lebowitz}(2004)}]{goldstein2004boltzmann}
\bibinfo{author}{\bibfnamefont{S.}~\bibnamefont{Goldstein}} \bibnamefont{and}
  \bibinfo{author}{\bibfnamefont{J.~L.} \bibnamefont{Lebowitz}},
  \bibinfo{journal}{\href{https://www.sciencedirect.com/science/article/pii/S0167278904000211}{Physica
  D}} \textbf{\bibinfo{volume}{193}}, \bibinfo{pages}{53}
  (\bibinfo{year}{2004}).

\bibitem[{\citenamefont{Brush}(2004)}]{brush2004history}
\bibinfo{author}{\bibfnamefont{S.~G.} \bibnamefont{Brush}},
  \emph{\bibinfo{title}{History of the Kinetic Theory of Gases}}
  (\bibinfo{publisher}{Istituto della Enciclopedia Italiana},
  \bibinfo{year}{2004}), vol.~\bibinfo{volume}{7}, chap.~\bibinfo{chapter}{44}.

\bibitem[{\citenamefont{Evans et~al.}(1993)\citenamefont{Evans, Cohen, and
  Morriss}}]{evans1993probability}
\bibinfo{author}{\bibfnamefont{D.~J.} \bibnamefont{Evans}},
  \bibinfo{author}{\bibfnamefont{E.~G.~D.} \bibnamefont{Cohen}},
  \bibnamefont{and} \bibinfo{author}{\bibfnamefont{G.~P.}
  \bibnamefont{Morriss}},
  \bibinfo{journal}{\href{https://journals.aps.org/prl/abstract/10.1103/PhysRevLett.71.2401}{Phys.
  Rev. Lett.}} \textbf{\bibinfo{volume}{71}}, \bibinfo{pages}{2401}
  (\bibinfo{year}{1993}).

\bibitem[{\citenamefont{Jarzynski}(1997)}]{jarzynski1997nonequilibrium}
\bibinfo{author}{\bibfnamefont{C.}~\bibnamefont{Jarzynski}},
  \bibinfo{journal}{\href{https://journals.aps.org/prl/abstract/10.1103/PhysRevLett.78.2690}{Phys.
  Rev. Lett.}} \textbf{\bibinfo{volume}{78}}, \bibinfo{pages}{2690}
  (\bibinfo{year}{1997}).

\bibitem[{\citenamefont{Crooks}(1999)}]{crooks1999entropy}
\bibinfo{author}{\bibfnamefont{G.~E.} \bibnamefont{Crooks}},
  \bibinfo{journal}{\href{https://journals.aps.org/pre/abstract/10.1103/PhysRevE.60.2721}{Phys.
  Rev. E}} \textbf{\bibinfo{volume}{60}}, \bibinfo{pages}{2721}
  (\bibinfo{year}{1999}).

\bibitem[{\citenamefont{Seifert}(2012)}]{seifert2012stochastic}
\bibinfo{author}{\bibfnamefont{U.}~\bibnamefont{Seifert}},
  \bibinfo{journal}{\href{https://iopscience.iop.org/article/10.1088/0034-4885/75/12/126001}{Rep.
  Prog. Phys.}} \textbf{\bibinfo{volume}{75}}, \bibinfo{pages}{126001}
  (\bibinfo{year}{2012}).

\bibitem[{\citenamefont{Luposchainsky et~al.}(2013)\citenamefont{Luposchainsky,
  Barato, and Hinrichsen}}]{luposchainsky2013strong}
\bibinfo{author}{\bibfnamefont{D.}~\bibnamefont{Luposchainsky}},
  \bibinfo{author}{\bibfnamefont{A.~C.} \bibnamefont{Barato}},
  \bibnamefont{and}
  \bibinfo{author}{\bibfnamefont{H.}~\bibnamefont{Hinrichsen}},
  \bibinfo{journal}{\href{https://www.researchgate.net/publication/236913916_Strong_fluctuation_theorem_for_nonstationary_nonequilibrium_systems}{Phys.
  Rev. E}} \textbf{\bibinfo{volume}{87}}, \bibinfo{pages}{042108}
  (\bibinfo{year}{2013}).

\bibitem[{\citenamefont{Talkner et~al.}(2008)\citenamefont{Talkner, H{\"a}nggi,
  and Morillo}}]{talkner2008microcanonical}
\bibinfo{author}{\bibfnamefont{P.}~\bibnamefont{Talkner}},
  \bibinfo{author}{\bibfnamefont{P.}~\bibnamefont{H{\"a}nggi}},
  \bibnamefont{and} \bibinfo{author}{\bibfnamefont{M.}~\bibnamefont{Morillo}},
  \bibinfo{journal}{\href{https://journals.aps.org/pre/abstract/10.1103/PhysRevE.77.051131}{Phys.
  Rev. E}} \textbf{\bibinfo{volume}{77}}, \bibinfo{pages}{051131}
  (\bibinfo{year}{2008}).

\bibitem[{\citenamefont{Esposito et~al.}(2009)\citenamefont{Esposito, Harbola,
  and Mukamel}}]{esposito2009nonequilibrium}
\bibinfo{author}{\bibfnamefont{M.}~\bibnamefont{Esposito}},
  \bibinfo{author}{\bibfnamefont{U.}~\bibnamefont{Harbola}}, \bibnamefont{and}
  \bibinfo{author}{\bibfnamefont{S.}~\bibnamefont{Mukamel}},
  \bibinfo{journal}{\href{https://journals.aps.org/rmp/abstract/10.1103/RevModPhys.81.1665}{Rev.
  Mod. Phys.}} \textbf{\bibinfo{volume}{81}}, \bibinfo{pages}{1665}
  (\bibinfo{year}{2009}).

\bibitem[{\citenamefont{Rana et~al.}(2013)\citenamefont{Rana, Lahiri, and
  Jayannavar}}]{rana2013generalized}
\bibinfo{author}{\bibfnamefont{S.}~\bibnamefont{Rana}},
  \bibinfo{author}{\bibfnamefont{S.}~\bibnamefont{Lahiri}}, \bibnamefont{and}
  \bibinfo{author}{\bibfnamefont{A.}~\bibnamefont{Jayannavar}},
  \bibinfo{journal}{\href{https://www.ias.ac.in/article/fulltext/pram/080/02/0207-0222}{Pramana
  J. Phys.}} \textbf{\bibinfo{volume}{80}}, \bibinfo{pages}{207}
  (\bibinfo{year}{2013}).

\bibitem[{\citenamefont{Esposito and Mukamel}(2006)}]{esposito2006fluctuation}
\bibinfo{author}{\bibfnamefont{M.}~\bibnamefont{Esposito}} \bibnamefont{and}
  \bibinfo{author}{\bibfnamefont{S.}~\bibnamefont{Mukamel}},
  \bibinfo{journal}{\href{https://journals.aps.org/pre/abstract/10.1103/PhysRevE.73.046129}{Phys.
  Rev. E}} \textbf{\bibinfo{volume}{73}}, \bibinfo{pages}{046129}
  (\bibinfo{year}{2006}).

\bibitem[{\citenamefont{Brunelli et~al.}(2018)\citenamefont{Brunelli, Fusco,
  Landig, Wieczorek, Hoelscher-Obermaier, Landi, Semi{\~a}o, Ferraro, Kiesel,
  Donner et~al.}}]{brunelli2018experimental}
\bibinfo{author}{\bibfnamefont{M.}~\bibnamefont{Brunelli}},
  \bibinfo{author}{\bibfnamefont{L.}~\bibnamefont{Fusco}},
  \bibinfo{author}{\bibfnamefont{R.}~\bibnamefont{Landig}},
  \bibinfo{author}{\bibfnamefont{W.}~\bibnamefont{Wieczorek}},
  \bibinfo{author}{\bibfnamefont{J.}~\bibnamefont{Hoelscher-Obermaier}},
  \bibinfo{author}{\bibfnamefont{G.}~\bibnamefont{Landi}},
  \bibinfo{author}{\bibfnamefont{F.~L.} \bibnamefont{Semi{\~a}o}},
  \bibinfo{author}{\bibfnamefont{A.}~\bibnamefont{Ferraro}},
  \bibinfo{author}{\bibfnamefont{N.}~\bibnamefont{Kiesel}},
  \bibinfo{author}{\bibfnamefont{T.}~\bibnamefont{Donner}},
  \bibnamefont{et~al.},
  \bibinfo{journal}{\href{https://journals.aps.org/prl/abstract/10.1103/PhysRevLett.121.160604}{Phys.
  Rev. Lett.}} \textbf{\bibinfo{volume}{121}}, \bibinfo{pages}{160604}
  (\bibinfo{year}{2018}).

\bibitem[{\citenamefont{Manzano et~al.}(2018)\citenamefont{Manzano, Horowitz,
  and Parrondo}}]{manzano2018quantum}
\bibinfo{author}{\bibfnamefont{G.}~\bibnamefont{Manzano}},
  \bibinfo{author}{\bibfnamefont{J.~M.} \bibnamefont{Horowitz}},
  \bibnamefont{and} \bibinfo{author}{\bibfnamefont{J.~M.~R.}
  \bibnamefont{Parrondo}},
  \bibinfo{journal}{\href{https://journals.aps.org/prx/abstract/10.1103/PhysRevX.8.031037}{Phys.
  Rev. X}} \textbf{\bibinfo{volume}{8}}, \bibinfo{pages}{031037}
  (\bibinfo{year}{2018}).

\bibitem[{\citenamefont{Popescu et~al.}(2006)\citenamefont{Popescu, Short, and
  Winter}}]{popescu2006entanglement}
\bibinfo{author}{\bibfnamefont{S.}~\bibnamefont{Popescu}},
  \bibinfo{author}{\bibfnamefont{A.~J.} \bibnamefont{Short}}, \bibnamefont{and}
  \bibinfo{author}{\bibfnamefont{A.}~\bibnamefont{Winter}},
  \bibinfo{journal}{\href{https://www.nature.com/articles/nphys444}{Nat.
  Phys.}} \textbf{\bibinfo{volume}{2}}, \bibinfo{pages}{754}
  (\bibinfo{year}{2006}).

\bibitem[{\citenamefont{Deutsch}(2010)}]{deutsch2010thermodynamic}
\bibinfo{author}{\bibfnamefont{J.~M.} \bibnamefont{Deutsch}},
  \bibinfo{journal}{\href{https://iopscience.iop.org/article/10.1088/1367-2630/12/7/075021/meta}{New
  J. Phys.}} \textbf{\bibinfo{volume}{12}}, \bibinfo{pages}{075021}
  (\bibinfo{year}{2010}).

\bibitem[{\citenamefont{Santos et~al.}(2012)\citenamefont{Santos, Polkovnikov,
  and Rigol}}]{santos2012weak}
\bibinfo{author}{\bibfnamefont{L.~F.} \bibnamefont{Santos}},
  \bibinfo{author}{\bibfnamefont{A.}~\bibnamefont{Polkovnikov}},
  \bibnamefont{and} \bibinfo{author}{\bibfnamefont{M.}~\bibnamefont{Rigol}},
  \bibinfo{journal}{\href{https://journals.aps.org/pre/abstract/10.1103/PhysRevE.86.010102}{Phys.
  Rev. E}} \textbf{\bibinfo{volume}{86}}, \bibinfo{pages}{010102(R)}
  (\bibinfo{year}{2012}).

\bibitem[{\citenamefont{Deutsch et~al.}(2020)\citenamefont{Deutsch,
  {\v{S}}afr{\'a}nek, and Aguirre}}]{deutsch2020probabilistic}
\bibinfo{author}{\bibfnamefont{J.~M.} \bibnamefont{Deutsch}},
  \bibinfo{author}{\bibfnamefont{D.}~\bibnamefont{{\v{S}}afr{\'a}nek}},
  \bibnamefont{and} \bibinfo{author}{\bibfnamefont{A.}~\bibnamefont{Aguirre}},
  \bibinfo{journal}{\href{https://journals.aps.org/pre/abstract/10.1103/PhysRevE.101.032112}{Phys.
  Rev. E}} \textbf{\bibinfo{volume}{101}}, \bibinfo{pages}{032112}
  (\bibinfo{year}{2020}).

\bibitem[{\citenamefont{Coffman et~al.}(2000)\citenamefont{Coffman, Kundu, and
  Wootters}}]{coffman2000distributed}
\bibinfo{author}{\bibfnamefont{V.}~\bibnamefont{Coffman}},
  \bibinfo{author}{\bibfnamefont{J.}~\bibnamefont{Kundu}}, \bibnamefont{and}
  \bibinfo{author}{\bibfnamefont{W.~K.} \bibnamefont{Wootters}},
  \bibinfo{journal}{\href{https://journals.aps.org/pra/abstract/10.1103/PhysRevA.61.052306}{Phys.
  Rev. A}} \textbf{\bibinfo{volume}{61}}, \bibinfo{pages}{052306}
  (\bibinfo{year}{2000}).

\bibitem[{\citenamefont{Rigolin et~al.}(2006)\citenamefont{Rigolin,
  de~Oliveira, and de~Oliveira}}]{rigolin2006operational}
\bibinfo{author}{\bibfnamefont{G.}~\bibnamefont{Rigolin}},
  \bibinfo{author}{\bibfnamefont{T.~R.} \bibnamefont{de~Oliveira}},
  \bibnamefont{and} \bibinfo{author}{\bibfnamefont{M.~C.}
  \bibnamefont{de~Oliveira}},
  \bibinfo{journal}{\href{https://journals.aps.org/pra/abstract/10.1103/PhysRevA.74.022314}{Phys.
  Rev. A}} \textbf{\bibinfo{volume}{74}}, \bibinfo{pages}{022314}
  (\bibinfo{year}{2006}).

\bibitem[{\citenamefont{Das et~al.}(2008)\citenamefont{Das, Shankaranarayanan,
  and Sur}}]{das2008black}
\bibinfo{author}{\bibfnamefont{S.}~\bibnamefont{Das}},
  \bibinfo{author}{\bibfnamefont{S.}~\bibnamefont{Shankaranarayanan}},
  \bibnamefont{and} \bibinfo{author}{\bibfnamefont{S.}~\bibnamefont{Sur}},
  \bibinfo{journal}{\href{https://arxiv.org/abs/0806.0402}{arXiv preprint
  arXiv:0806.0402} [gr-qc]}  (\bibinfo{year}{2008}).

\bibitem[{\citenamefont{Bennett et~al.}(1993)\citenamefont{Bennett, Brassard,
  Cr{\'e}peau, Jozsa, Peres, and Wootters}}]{bennett1993teleporting}
\bibinfo{author}{\bibfnamefont{C.~H.} \bibnamefont{Bennett}},
  \bibinfo{author}{\bibfnamefont{G.}~\bibnamefont{Brassard}},
  \bibinfo{author}{\bibfnamefont{C.}~\bibnamefont{Cr{\'e}peau}},
  \bibinfo{author}{\bibfnamefont{R.}~\bibnamefont{Jozsa}},
  \bibinfo{author}{\bibfnamefont{A.}~\bibnamefont{Peres}}, \bibnamefont{and}
  \bibinfo{author}{\bibfnamefont{W.~K.} \bibnamefont{Wootters}},
  \bibinfo{journal}{\href{https://journals.aps.org/prl/abstract/10.1103/PhysRevLett.70.1895}{Phys.
  Rev. Lett.}} \textbf{\bibinfo{volume}{70}}, \bibinfo{pages}{1895}
  (\bibinfo{year}{1993}).

\bibitem[{\citenamefont{Helwig et~al.}(2012)\citenamefont{Helwig, Cui, Latorre,
  Riera, and Lo}}]{helwig2012absolute}
\bibinfo{author}{\bibfnamefont{W.}~\bibnamefont{Helwig}},
  \bibinfo{author}{\bibfnamefont{W.}~\bibnamefont{Cui}},
  \bibinfo{author}{\bibfnamefont{J.~I.} \bibnamefont{Latorre}},
  \bibinfo{author}{\bibfnamefont{A.}~\bibnamefont{Riera}}, \bibnamefont{and}
  \bibinfo{author}{\bibfnamefont{H.-K.} \bibnamefont{Lo}},
  \bibinfo{journal}{\href{https://journals.aps.org/pra/abstract/10.1103/PhysRevA.86.052335}{Phys.
  Rev. A}} \textbf{\bibinfo{volume}{86}}, \bibinfo{pages}{052335}
  (\bibinfo{year}{2012}).

\bibitem[{\citenamefont{Zhang et~al.}(2011)\citenamefont{Zhang, Grover, and
  Vishwanath}}]{zhang2011entanglement}
\bibinfo{author}{\bibfnamefont{Y.}~\bibnamefont{Zhang}},
  \bibinfo{author}{\bibfnamefont{T.}~\bibnamefont{Grover}}, \bibnamefont{and}
  \bibinfo{author}{\bibfnamefont{A.}~\bibnamefont{Vishwanath}},
  \bibinfo{journal}{\href{https://journals.aps.org/prl/abstract/10.1103/PhysRevLett.107.067202}{Phys.
  Rev. Lett.}} \textbf{\bibinfo{volume}{107}}, \bibinfo{pages}{067202}
  (\bibinfo{year}{2011}).

\bibitem[{\citenamefont{Jiang et~al.}(2012)\citenamefont{Jiang, Wang, and
  Balents}}]{jiang2012identifying}
\bibinfo{author}{\bibfnamefont{H.-C.} \bibnamefont{Jiang}},
  \bibinfo{author}{\bibfnamefont{Z.}~\bibnamefont{Wang}}, \bibnamefont{and}
  \bibinfo{author}{\bibfnamefont{L.}~\bibnamefont{Balents}},
  \bibinfo{journal}{\href{https://www.nature.com/articles/nphys2465}{Nat.
  Phys.}} \textbf{\bibinfo{volume}{8}}, \bibinfo{pages}{902}
  (\bibinfo{year}{2012}).

\bibitem[{\citenamefont{Sugiura and Shimizu}(2012)}]{sugiura2012thermal}
\bibinfo{author}{\bibfnamefont{S.}~\bibnamefont{Sugiura}} \bibnamefont{and}
  \bibinfo{author}{\bibfnamefont{A.}~\bibnamefont{Shimizu}},
  \bibinfo{journal}{\href{https://journals.aps.org/prl/abstract/10.1103/PhysRevLett.108.240401}{Phys.
  Rev. Lett.}} \textbf{\bibinfo{volume}{108}}, \bibinfo{pages}{240401}
  (\bibinfo{year}{2012}).

\bibitem[{\citenamefont{Sugiura and Shimizu}(2013)}]{sugiura2013canonical}
\bibinfo{author}{\bibfnamefont{S.}~\bibnamefont{Sugiura}} \bibnamefont{and}
  \bibinfo{author}{\bibfnamefont{A.}~\bibnamefont{Shimizu}},
  \bibinfo{journal}{\href{https://journals.aps.org/prl/abstract/10.1103/PhysRevLett.111.010401}{Phys.
  Rev. Lett.}} \textbf{\bibinfo{volume}{111}}, \bibinfo{pages}{010401}
  (\bibinfo{year}{2013}).

\bibitem[{\citenamefont{Nakagawa et~al.}(2018)\citenamefont{Nakagawa, Watanabe,
  Fujita, and Sugiura}}]{nakagawa2018universality}
\bibinfo{author}{\bibfnamefont{Y.~O.} \bibnamefont{Nakagawa}},
  \bibinfo{author}{\bibfnamefont{M.}~\bibnamefont{Watanabe}},
  \bibinfo{author}{\bibfnamefont{H.}~\bibnamefont{Fujita}}, \bibnamefont{and}
  \bibinfo{author}{\bibfnamefont{S.}~\bibnamefont{Sugiura}},
  \bibinfo{journal}{\href{https://www.nature.com/articles/s41467-018-03883-9}{Nat.
  Comm.}} \textbf{\bibinfo{volume}{9}}, \bibinfo{pages}{1635}
  (\bibinfo{year}{2018}).

\bibitem[{\citenamefont{Nelder and Mead}(1965)}]{nelder1965simplex}
\bibinfo{author}{\bibfnamefont{J.~A.} \bibnamefont{Nelder}} \bibnamefont{and}
  \bibinfo{author}{\bibfnamefont{R.}~\bibnamefont{Mead}},
  \bibinfo{journal}{\href{https://journals.aps.org/rmp/abstract/10.1103/RevModPhys.82.277}{Comput.
  J.}} \textbf{\bibinfo{volume}{7}}, \bibinfo{pages}{308}
  (\bibinfo{year}{1965}).

\bibitem[{\citenamefont{Faiez and {\v{S}}afr{\'a}nek}(2020)}]{faiez2020much}
\bibinfo{author}{\bibfnamefont{D.}~\bibnamefont{Faiez}} \bibnamefont{and}
  \bibinfo{author}{\bibfnamefont{D.}~\bibnamefont{{\v{S}}afr{\'a}nek}},
  \bibinfo{journal}{\href{https://journals.aps.org/prb/abstract/10.1103/PhysRevB.101.060401}{Phys.
  Rev. B}} \textbf{\bibinfo{volume}{101}}, \bibinfo{pages}{060401}
  (\bibinfo{year}{2020}).

\bibitem[{\citenamefont{Yamaguchi}(2019)}]{yamaguchi2018theorems}
\bibinfo{author}{\bibfnamefont{K.}~\bibnamefont{Yamaguchi}},
  \bibinfo{journal}{\href{https://journals.jps.jp/doi/10.7566/JPSJ.88.064002}{J.
  Phys. Soc. Jpn.}} \textbf{\bibinfo{volume}{88}}, \bibinfo{pages}{064002}
  (\bibinfo{year}{2019}).

\bibitem[{\citenamefont{S{\o}rensen et~al.}(2007)\citenamefont{S{\o}rensen,
  Chang, Laflorencie, and Affleck}}]{sorensen2007quantum}
\bibinfo{author}{\bibfnamefont{E.~S.} \bibnamefont{S{\o}rensen}},
  \bibinfo{author}{\bibfnamefont{M.-S.} \bibnamefont{Chang}},
  \bibinfo{author}{\bibfnamefont{N.}~\bibnamefont{Laflorencie}},
  \bibnamefont{and} \bibinfo{author}{\bibfnamefont{I.}~\bibnamefont{Affleck}},
  \bibinfo{journal}{\href{https://iopscience.iop.org/article/10.1088/1742-5468/2007/08/P08003/meta}{J.
  Stat. Mech.}} \textbf{\bibinfo{volume}{2007}}, \bibinfo{pages}{P08003}
  (\bibinfo{year}{2007}).

\bibitem[{\citenamefont{H{\"a}nggi et~al.}(2008)\citenamefont{H{\"a}nggi,
  Ingold, and Talkner}}]{hanggi2008finite}
\bibinfo{author}{\bibfnamefont{P.}~\bibnamefont{H{\"a}nggi}},
  \bibinfo{author}{\bibfnamefont{G.-L.} \bibnamefont{Ingold}},
  \bibnamefont{and} \bibinfo{author}{\bibfnamefont{P.}~\bibnamefont{Talkner}},
  \bibinfo{journal}{\href{https://iopscience.iop.org/article/10.1088/1367-2630/10/11/115008/meta}{New
  J. Phys.}} \textbf{\bibinfo{volume}{10}}, \bibinfo{pages}{115008}
  (\bibinfo{year}{2008}).

\bibitem[{\citenamefont{H{\"o}rhammer and
  B{\"u}ttner}(2008)}]{horhammer2008information}
\bibinfo{author}{\bibfnamefont{C.}~\bibnamefont{H{\"o}rhammer}}
  \bibnamefont{and}
  \bibinfo{author}{\bibfnamefont{H.}~\bibnamefont{B{\"u}ttner}},
  \bibinfo{journal}{\href{https://link.springer.com/article/10.1007/s10955-008-9640-x}{J.
  Stat. Phys.}} \textbf{\bibinfo{volume}{133}}, \bibinfo{pages}{1161}
  (\bibinfo{year}{2008}).

\bibitem[{\citenamefont{Eisert et~al.}(2010)\citenamefont{Eisert, Cramer, and
  Plenio}}]{eisert2008area}
\bibinfo{author}{\bibfnamefont{J.}~\bibnamefont{Eisert}},
  \bibinfo{author}{\bibfnamefont{M.}~\bibnamefont{Cramer}}, \bibnamefont{and}
  \bibinfo{author}{\bibfnamefont{M.~B.} \bibnamefont{Plenio}},
  \bibinfo{journal}{\href{https://journals.aps.org/rmp/abstract/10.1103/RevModPhys.82.277}{Rev.
  Mod. Phys.}} \textbf{\bibinfo{volume}{82}}, \bibinfo{pages}{277}
  (\bibinfo{year}{2010}).

\bibitem[{\citenamefont{Laflorencie}(2016)}]{laflorencie2016quantum}
\bibinfo{author}{\bibfnamefont{N.}~\bibnamefont{Laflorencie}},
  \bibinfo{journal}{\href{https://www.sciencedirect.com/science/article/abs/pii/S0370157316301582}{Phys.
  Rep.}} \textbf{\bibinfo{volume}{646}}, \bibinfo{pages}{1}
  (\bibinfo{year}{2016}).

\bibitem[{\citenamefont{Davies}(1977)}]{davies1977physics}
\bibinfo{author}{\bibfnamefont{P.~C.~W.} \bibnamefont{Davies}},
  \emph{\bibinfo{title}{The Physics of Time Asymmetry}}
  (\bibinfo{publisher}{University of California Press},
  \bibinfo{address}{Oakland}, \bibinfo{year}{1977}).

\bibitem[{\citenamefont{Aguirre and Gratton}(2003)}]{aguirre2003inflation}
\bibinfo{author}{\bibfnamefont{A.}~\bibnamefont{Aguirre}} \bibnamefont{and}
  \bibinfo{author}{\bibfnamefont{S.}~\bibnamefont{Gratton}},
  \bibinfo{journal}{\href{https://journals.aps.org/prd/abstract/10.1103/PhysRevD.67.083515}{Phys.
  Rev. D}} \textbf{\bibinfo{volume}{67}}, \bibinfo{pages}{083515}
  (\bibinfo{year}{2003}).

\bibitem[{\citenamefont{Albrecht and Sorbo}(2004)}]{albrecht2004can}
\bibinfo{author}{\bibfnamefont{A.}~\bibnamefont{Albrecht}} \bibnamefont{and}
  \bibinfo{author}{\bibfnamefont{L.}~\bibnamefont{Sorbo}},
  \bibinfo{journal}{\href{https://journals.aps.org/prd/abstract/10.1103/PhysRevD.70.063528}{Phys.
  Rev. D}} \textbf{\bibinfo{volume}{70}}, \bibinfo{pages}{063528}
  (\bibinfo{year}{2004}).

\bibitem[{\citenamefont{Penrose}(2004)}]{penrose2004road}
\bibinfo{author}{\bibfnamefont{R.}~\bibnamefont{Penrose}},
  \emph{\bibinfo{title}{The Road to Reality. A Complete Guide to the Laws of
  the Universe, Joanthan Cape}} (\bibinfo{publisher}{Random House, London},
  \bibinfo{year}{2004}).

\bibitem[{\citenamefont{Carroll and Chen}(2004)}]{carroll2004spontaneous}
\bibinfo{author}{\bibfnamefont{S.~M.} \bibnamefont{Carroll}} \bibnamefont{and}
  \bibinfo{author}{\bibfnamefont{J.}~\bibnamefont{Chen}},
  \bibinfo{journal}{\href{https://arxiv.org/abs/hep-th/0410270}{arXiv:hep-th/0410270}
  [hep-th]}  (\bibinfo{year}{2004}).

\bibitem[{\citenamefont{Aguirre et~al.}(2012)\citenamefont{Aguirre, Carroll,
  and Johnson}}]{aguirre2012out}
\bibinfo{author}{\bibfnamefont{A.}~\bibnamefont{Aguirre}},
  \bibinfo{author}{\bibfnamefont{S.~M.} \bibnamefont{Carroll}},
  \bibnamefont{and} \bibinfo{author}{\bibfnamefont{M.~C.}
  \bibnamefont{Johnson}},
  \bibinfo{journal}{\href{https://iopscience.iop.org/article/10.1088/1475-7516/2012/02/024/meta}{J.
  Cosmol. Astropart. Phys.}} \textbf{\bibinfo{volume}{2012}},
  \bibinfo{pages}{024} (\bibinfo{year}{2012}).

\end{thebibliography}
\end{document}